\documentclass[a4paper,11pt]{article}
\usepackage{jinstpub} 
\usepackage{lineno}
\usepackage{siunitx}
\usepackage{booktabs}
\usepackage{float}
\usepackage{booktabs}
\usepackage{siunitx}
\usepackage{float}
\usepackage{subcaption}

\title{\boldmath The Cryogenic System of DMRadio-50L}

\collaboration{DMRadio Collaboration}
\author[a,b]{V.~Ankel,}
\author[c,b]{C.~Bartram,}
\author[d]{J.~Begin,}
\author[a,b]{C.~Bell,}
\author[d]{S.~Chaudhuri,}
\author[c,b]{H.-M.~Cho,}
\author[a,b]{J.~Corbin,}
\author[c]{W.~Craddock,}
\author[e]{S.~Cuadra,}
\author[a,b,c]{A.~Droster,}
\author[f]{J.~Echevers,}
\author[g]{E.~Engelhardt,}
\author[e]{J.~T.~Fry,}
\author[d]{J.~Fu,}
\author[a,b,c]{K.~D.~Irwin,}
\author[a,b]{A.~Keller,}
\author[d]{R.~Kolevatov,}
\author[a,b]{A.~Kunder,}
\author[c,b]{D.~Li,}
\author[a,b]{M.~Marangola,}
\author[d]{N.~Otto,}
\author[e]{K.~M.~W.~Pappas,}
\author[e]{E.~Pariset,}
\author[h]{S.~Puranam,}
\author[i,j]{P.~Quassolo,}
\author[a,b]{N.~M.~Rapidis,}
\author[h,g]{C.~P.~Salemi,}
\author[a,b]{M.~Simanovskaia,}
\author[a,b]{J.~Singh,}
\author[a,c,b]{P.~Stark,}
\author[a,b]{E.~C.~van~Assendelft,}
\author[f,g]{K.~van~Bibber,}
\author[e]{K.~J.~Vetter,}
\author[a,b]{K.~Wells,}
\author[d]{J.~Wiedemann,}
\author[e]{L.~Winslow,}
\author[a,b]{D.~Wright,}
\author[c,b]{A.~K.~Yi,}
\author[a,c,b]{and B.~F.~Zemenu}

\affiliation[a]{Department of Physics, Stanford University, Stanford, CA 94305}
\affiliation[b]{Kavli Institute for Particle Astrophysics and Cosmology, Stanford University, Stanford, CA 94305}
\affiliation[c]{SLAC Accelerator National Laboratory, Menlo Park, CA 94025}
\affiliation[d]{Department of Physics, Princeton University, Princeton, NJ 08544}
\affiliation[e]{Laboratory of Nuclear Science, Massachusetts Institute of Technology, Cambridge, MA 02139}
\affiliation[f]{Department of Nuclear Engineering, University of California, Berkeley, Berkeley, CA 94720}
\affiliation[g]{Physics Division, Lawrence Berkeley National Laboratory, Berkeley, CA 94720}
\affiliation[h]{Department of Physics, University of California, Berkeley, Berkeley, CA 94720}
\affiliation[i]{Applied Science and Technology, College of Engineering, University of California, Berkeley, Berkeley, CA 94720}
\affiliation[j]{Accelerator Technology and Applied Physics Division, Lawrence Berkeley National Laboratory, Berkeley, CA 94720}

\emailAdd{ayakeller@stanford.edu}
\emailAdd{simanovskaia@stanford.edu}

\abstract{
The DMRadio-50L experiment is designed to search for axion dark matter in the \SI{5}{\kilo\hertz} -- \SI{5}{\mega\hertz} frequency range using a lumped-element LC resonator and a toroidal magnet and to serve as a testbed for quantum sensors. This paper describes the custom cryogenic system developed to meet the stringent requirements of the experiment within a standard laboratory environment. The system is designed to cool a \SI{200}{\kilo\gram} detector assembly to temperatures as low as \SI{50}{\milli\kelvin} while providing sufficient cooling power at multiple temperature stages. We present the conceptual design, technical implementation, and measured performance of the hybrid cryogenic system, which combines a horizontal dilution refrigerator with a large vertical payload cryostat.
}

\begin{document}
\maketitle
\flushbottom

\section{Introduction}

Axions are a compelling dark matter candidate which could simultaneously explain the absence of CP violation in QCD~\cite{Pec1977a,Pec1977b,Wei1978,Wil1978} and account for the observed dark matter abundance~\cite{Din1983,Pre1983,Abb1983,Teg2006,Her2008,Co2016,Gra2018,Tak2018}. Experimental searches for axion dark matter seek to detect a small narrowband electromagnetic power excess. To achieve sensitivity to such small signals, experimental setups must employ cutting-edge cryogenic systems that minimize thermal and readout noise. In axion searches using resonant detection techniques such as DMRadio-50L~\cite{50l_paper}, the signal-to-noise ratio benefits from lower temperatures, which lead to a reduction in both the thermal occupation of the resonator mode and the added noise of cryogenic amplifiers~\cite{Bro2022b,AlS2023}. Operation at mK temperatures is therefore essential for reaching the highest sensitivity.

The DMRadio-50L detector targets the \SI{5}{\kilo\hertz}--\SI{5}{\mega\hertz} resonance frequency range, corresponding to axion masses in the $\mathcal{O}(10^{-11}\text{--}10^{-8})$ eV regime. The experiment employs an inductor in combination with a tunable capacitor to form a lumped-element superconducting resonator with a high quality factor ($Q$). To suppress thermal noise and minimize resistive losses, both the resonator and its readout chain must be cooled to temperatures as low as possible subject to the available cooling power and heat loads. The system was therefore designed to achieve a base temperature below \SI{50}{mK} for the most sensitive components.

DMRadio-50L is also designed to serve as a testbed for quantum sensors such as the radio-frequency quantum upconverter (RQU)~\cite{Kue2024}. Such sensors have multiple applications, but they are essential for accelerating future axion searches: by enabling readout with noise below the standard quantum limit, they can substantially reduce the time required to scan a given range of axion masses. Their development benefits from a stable cryogenic platform that provides a low bath temperature, sufficient room for the implementation of test circuitry, low background magnetic fields, and access to a high-$Q$ superconducting resonator in the \si{\mega\hertz} frequency range. The cryogenic system described here is designed to meet these needs, establishing a versatile environment for developing and characterizing next-generation quantum sensors alongside the axion search.

The scale of the DMRadio-50L detector imposes substantial mechanical and thermal demands on the cryogenic system. The detector mass approaches \SI{200}{kg}, dominated by the superconducting magnet, shielding, superconducting structures, and support hardware with a total experimental volume of at least \SI{0.5}{\cubic\meter}. The system must therefore provide both large mechanical capacity and high cooling power across multiple temperature stages while preserving excellent thermal isolation to the mK stage housing the resonator and readout chain. In addition, because mechanical vibrations can couple into the readout chain as unwanted noise, effective vibration mitigation throughout the cryogenic and mechanical structures is essential to preserve detector stability and low-noise performance.

To address these challenges, we developed a hybrid cryogenic architecture combining a horizontal dilution refrigerator (DR) with a vertical payload cryostat -- an approach previously employed in SuperCDMS \cite{Hollister2017}. Section~\ref{sec:concept} describes the system requirements that drive the architecture of the cryogenic system, followed by a high-level overview of the design. Section~\ref{sec:technical} details the technical design of the cryogenic system, focusing on the various thermal paths, controls and readout, and seismic design considerations. In Section~\ref{sec:performance}, we report on the cryogenic performance of the cooling process along with the steady-state measurements. Section~\ref{sec:conclusion} concludes with the summary and outlook.
\section{Conceptual Design}
\label{sec:concept} 

\subsection{System Requirements}\label{sec:reqs}

The cryogenic system of DMRadio-50L must satisfy a set of thermal, mechanical, and operational requirements driven by both detector sensitivity and practical integration constraints. The principal requirements are summarized below.

\begin{enumerate}
    \item An experimental volume of at least \SI{0.5}{\cubic\meter}: The cryogenic system must accommodate the superconducting magnet, electromagnetic shielding, resonator components, structural supports, and associated readout hardware at or below \SI{4}{\kelvin}.
    \item External spatial constraints: The cryogenic system must be compatible with the physical constraints of the laboratory space in which it is installed. In particular, the system must be designed so that it can be assembled and operated with a laboratory ceiling height of 14 ft, while still allowing access for installation, maintenance, and overhead crane operations.
    \item Mechanical support of \SI{200}{\kilo\gram}: The second stage payload plate must be able to support the full detector mass of \SI{200}{\kilo\gram}. Of the \SI{200}{\kilo\gram}, \SI{65}{\kilo\gram} are supported by the still stage, and \SI{25}{\kilo\gram} are supported by the base stage.
    \item Cooling power of at least \SI{300}{\milli\watt} at \SI{3.3}{\kelvin} on the second stage payload plate: Adequate cooling capacity near \SI{4}{\kelvin} is required to maintain the superconducting magnet wound with niobium titanium (NbTi) wire in its persistent current mode. 
    \item Base temperature below \SI{50}{\milli\kelvin} on the base payload plate:  The base temperature stage contains the resonator and sensitive readout electronics attached to the base payload plate, which are cooled by cold fingers coming from the mixing chamber (MXC) plate and should be as cold as possible to minimize noise.
    \item Operating modes: The cryogenic platform must support multiple operating configurations to facilitate testing, commissioning, and maintenance. This includes the ability to pump all vacuum spaces to pressures below \SI{1e-4}{\milli\bar}, accommodate configurable wiring such as twisted-pair looms and coaxial lines, and provide connections for temperature sensors and heaters. In addition, the design must allow the DR and the payload cryostat to be cooled down and operated separately when necessary.
    \item Vibration isolation: Mechanical vibrations from pulse tube cryocoolers can degrade detector performance and introduce noise. The system must therefore incorporate mechanisms to mechanically isolate the pulse tubes from the payload to minimize vibration transmission to the payload and sensitive detector components.

\end{enumerate}

The DMRadio-50L cryogenic system is designed to satisfy these requirements while maintaining operational flexibility.

\begin{figure}[h]
    \centering
    \includegraphics[width=1\textwidth]{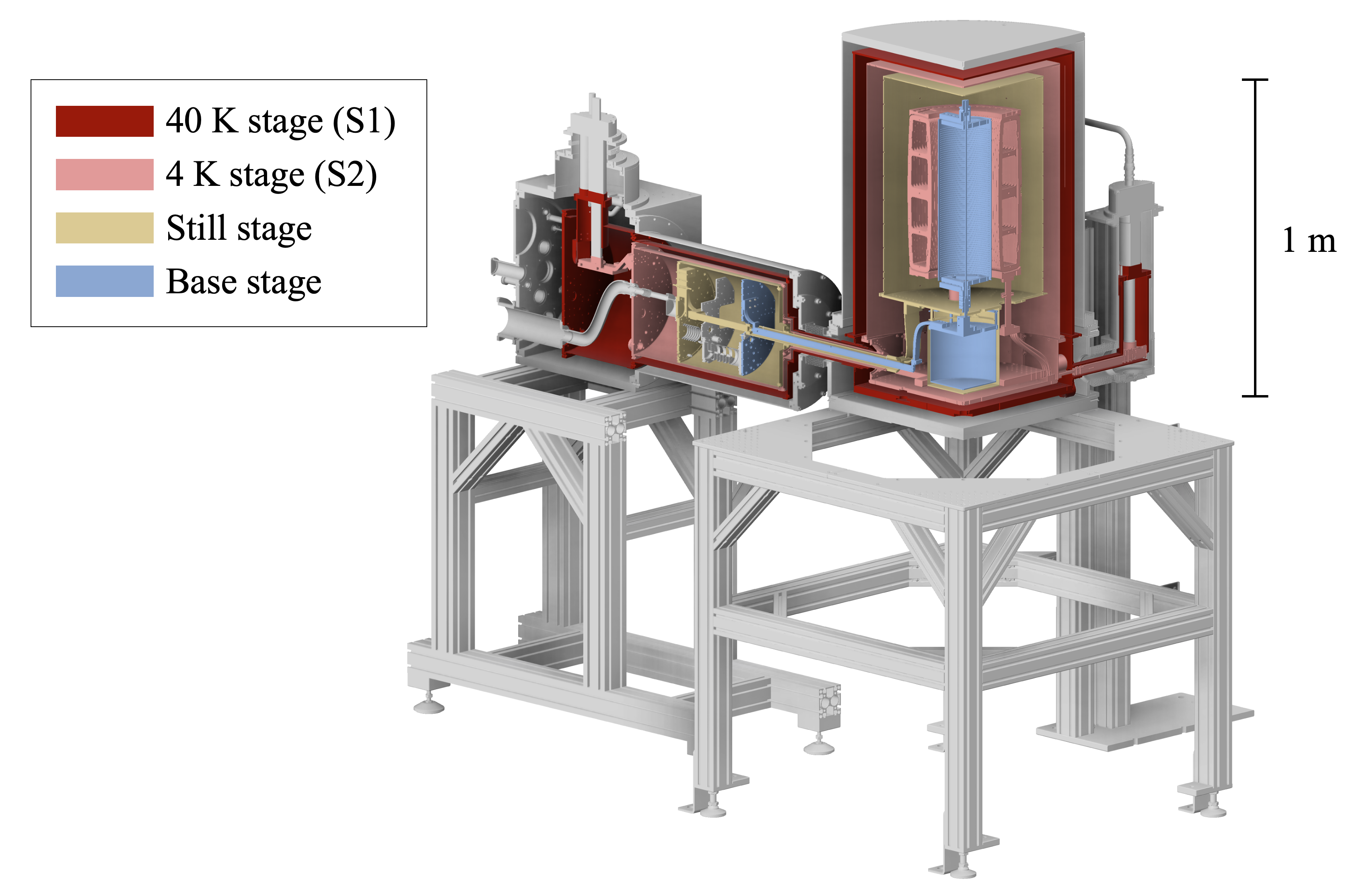} 
    \caption{DMRadio-50L cryogenic system CAD cross section with two perpendicular cutaways. The horizontal DR that provides cooling through cold fingers at all temperature stages is on the left, and the payload cryostat that houses the experimental volume with cold fingers coming from the first and second stages of a pulse tube cryocooler on the right.  Selected detector components such as the toroidal magnet and inductor are included to demonstrate the nested structure of the temperature stages. The base stage is shown in blue, the still stage in gold, the \SI{4}{\kelvin} stage (S2) in pink, and the \SI{40}{\kelvin} stage (S1) in red.}
    \label{fig:overview}
\end{figure}

\subsection{High-Level Overview}
\begin{figure*}[h!]
    \centering
    \includegraphics[width=1\textwidth]{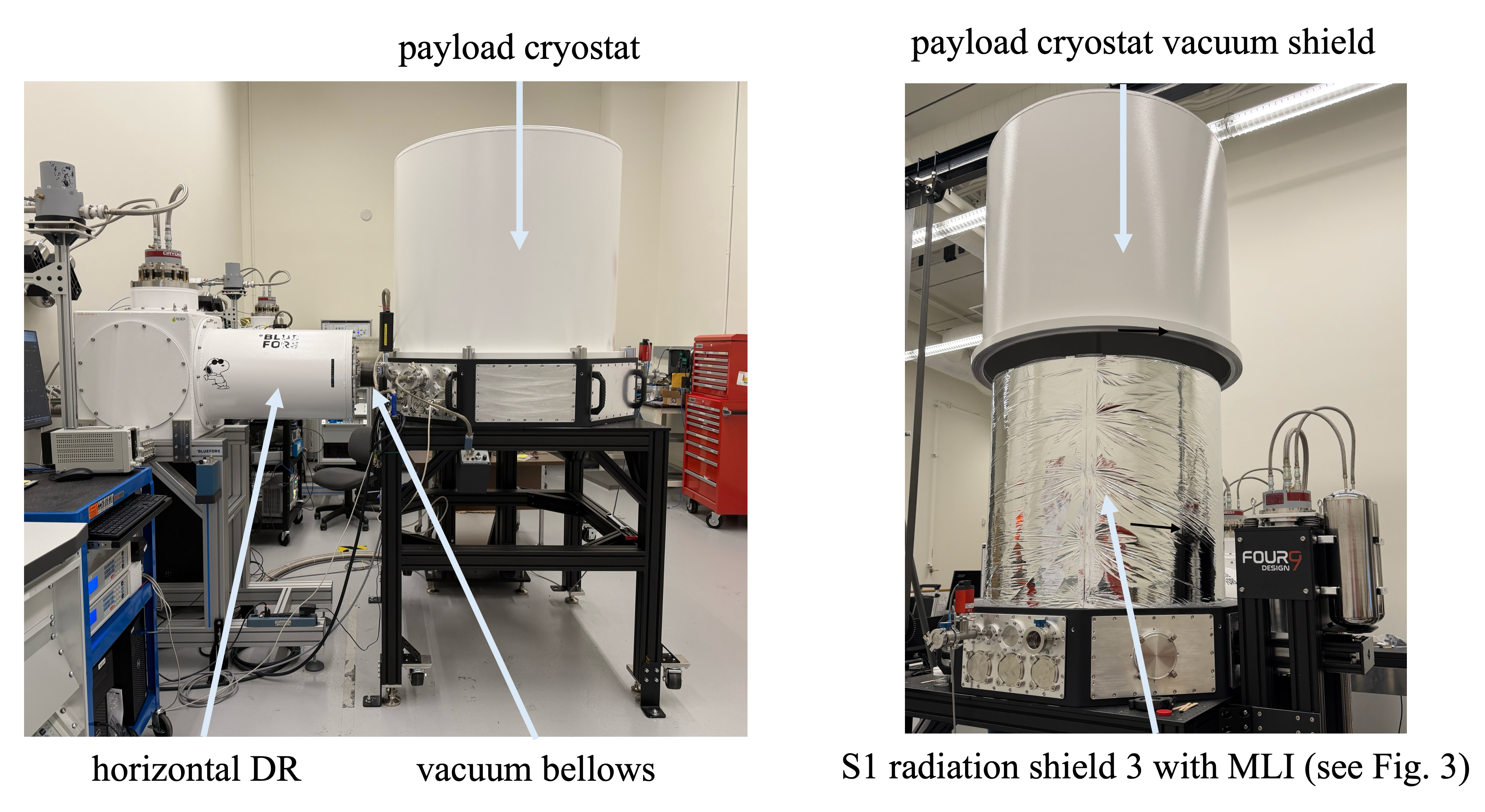}
    \caption{Photographs of the cryogenic system in the lab. The left image shows the fully closed-up and operating system with the horizontal DR and the payload cryostat connected via the vacuum bellows. The right image shows the payload cryostat vacuum shield being lowered above S1 radiation shield 3 with multi-layer insulation (MLI) during the assembly of the payload cryostat.}
    \label{fig:cryophoto}
\end{figure*}
The DMRadio-50L cryogenic system consists of two primary subsystems: a vertically oriented payload cryostat and a horizontally mounted DR. The two subsystems share a vacuum space and cool the payload components to the required temperatures. A CAD cross-section of the configuration is shown in Figure~\ref{fig:overview}, and photos of the assembled cryogenic system are shown in Figure~\ref{fig:cryophoto}. The payload cryostat is mounted on a rail system that allows it to translate away from the DR, enabling independent testing and maintenance operations. The interface between the two cryostats presents a multifaceted design challenge: it must selectively provide mechanical compliance or rigidity, enable high thermal conductance or strong thermal isolation, and simultaneously suppress radiative heat transfer between stages. 

The payload cryostat houses the experiment and provides cooling power at approximately \SI{40}{\kelvin} on the first stage (S1) and \SI{4}{\kelvin} on the second stage (S2) via a PT425-RM cryocooler \cite{Radebaugh2009}. The S2 payload plate supports a secondary aluminum (Al) ring at the same temperature, which mechanically supports the still and base payload plates and facilitates the assembly and disassembly of the lower-temperature systems. Feedthroughs on the side panels of the payload cryostat accommodate the required detector wiring.

The DR provides cooling power at the still and base temperature stages through cold finger assemblies that balance thermal conductance and mechanical compliance. A PT420-RM cryocooler is integrated into the DR to precool the still and MXC plates as well as the payload components thermally linked to the DR to the \SI{4}{\kelvin} stage temperature via heat switches, enabling efficient initial cooldown prior to DR operation. 

The two subsystems cooperate at the \SI{4}{\kelvin} stage to cool the majority of the payload thermal mass to approximately \SI{4}{\kelvin}. At the \SI{40}{\kelvin} stage, the two subsystems are thermally independent to avoid affecting DR performance by the payload heat loads.

Each temperature stage is surrounded by radiation shields anchored to the corresponding stage of the cryocooler. The magnet volume must be shielded from electromagnetic interference (EMI) with superconducting material, requiring a nested geometry in which the magnet is enclosed in a lower-temperature environment. Thermal links between all stages are designed to provide high conductance while accommodating differential thermal contraction. The estimated heat loads at each temperature stage are summarized in Table~\ref{tab:summary}. The design also accommodates future vibration isolation upgrades, including independent support structures for the S2 payload ring.


\begin{table}[h] 
\centering
\caption{Predicted steady-state heat loads at each temperature stage.}
\label{tab:summary}
\begin{tabular}{lc}
\hline
\textbf{Stage} & \textbf{Predicted heat load} \\
\hline
Base (MXC) & \SI{6}{\micro\watt} \\
Still & \SI{0.2}{\milli\watt} \\
\SI{4}{\kelvin} stage & \SI{.3}{\watt} \\
\SI{40}{\kelvin} stage & \SI{7}{\watt} \\
\hline
\end{tabular}
\label{tbl:heat_loads}
\end{table}

\section{Technical Design}
\label{sec:technical}

The DMRadio-50L cryogenic platform employs a multi-stage approach to distribute the thermal load according to the available cooling power at each temperature, allowing large intermediate heat loads to be absorbed at higher temperatures while reserving the limited dilution cooling capacity for only the most temperature-sensitive components.

Cooling to around \SI{4}{\kelvin} is achieved with two pulse tube cryocoolers. The custom payload cryostat designed and manufactured by Four9 Design\footnote{https://four9design.com/} incorporates a Cryomech PT425-RM pulse tube cryocooler, which is certified to provide \SI{50}{\watt} at \SI{45}{\kelvin}, \SI{2.35}{\watt} at \SI{4.2}{\kelvin}, and a base temperature of \SI{2.8}{\kelvin}. The second pulse tube cryocooler is integrated into the Bluefors\footnote{https://bluefors.com/} LH400 DR; the Cryomech PT420-RM pulse tube cryocooler is certified to provide \SI{50}{\watt} at \SI{45}{\kelvin}, \SI{1.8}{\watt} at \SI{4.2}{\kelvin}, and a base temperature of \SI{2.8}{\kelvin}. The payload components with colder base temperatures than \SI{4}{\kelvin} are first precooled to approximately \SI{4}{\kelvin} through heat switches thermally anchored to the second stage of the PT420-RM pulse tube cryocooler. This strategy reduces the heat that must later be removed by the DR and substantially shortens overall cooldown time.

Cooling below \SI{4}{\kelvin} is provided by the still and MXC of the Bluefors LH400 DR connected through cold fingers to the payload cryostat. The DR delivers over \SI{12}{\micro\watt} of cooling power at \SI{20}{\milli\kelvin} on the MXC plate and over \SI{10}{\milli\watt} at \SI{1}{\kelvin} on the still plate. The MXC plate base temperature is certified to be at most \SI{7}{\milli\kelvin}. This capacity is sufficient to cool the resonator and readout chain while maintaining operational margin against parasitic loads.

The following Sections describe the cooling paths for the relevant temperature stages. Additional details on component dimensions and thermal interfaces are provided in Appendix \ref{appendix:dimensions}. 


\subsection{Vacuum system}

The payload cryostat and the horizontal DR share a common vacuum space, connected by a soft vacuum bellows (labeled in Figure~\ref{fig:cryophoto}). When compressed to its installed length, this bellows has an inner diameter of \SI{76.2}{\milli\meter}, which constrains the allowable dimensions of the cold fingers and radiation shields passing between the two volumes. Consequently, the dimensions of all interconnecting thermal hardware are determined not only by thermal considerations but also by this geometric constraint. A second identical soft vacuum bellows connects the integrated PT425-RM cold head assembly to the main chamber of the payload cryostat. These bellows connections dictate the alignment tolerances and mechanical compliance requirements across subsystems. The narrow apertures accommodating the cold fingers to the DR and the PT425-RM cold head assembly can be seen on either side of the large vertical payload vacuum chamber in Figure~\ref{fig:overview}. The payload vacuum chamber, the integrated PT425-RM cold head assembly, and the horizontal DR are each independently supported by external mechanical frames. Structural loads are therefore carried by the frames rather than through the bellows connections, which serve as compliant vacuum interfaces rather than load-bearing elements.

The payload cryostat consists of an octagonal Al6061 base chamber capped by a cylindrical Al6061 vacuum chamber lid with an inner diameter of \SI{1}{\meter}, forming an internal payload vacuum chamber with an overall height of \SI{1}{\meter}. All major subsystem interfaces are consolidated on the base chamber panels, including the two bellows ports and electrical feedthroughs on KF25, KF50, and ISO63 flanges. 

\subsection{40 K stage (S1) thermal path \label{sec:1st_stage_path}}
\begin{figure*}[h!]
    \centering
    \includegraphics[width=1\textwidth]{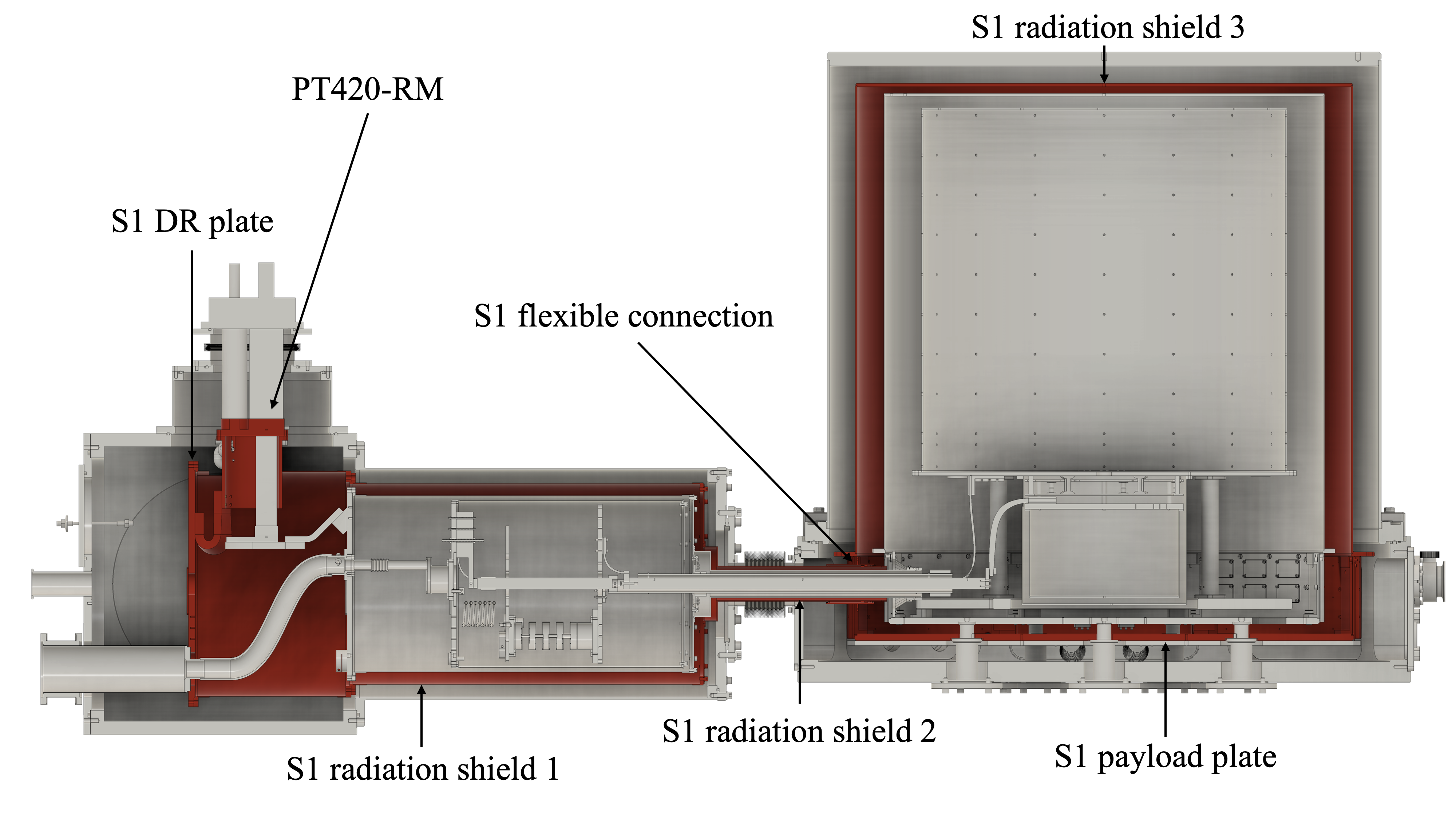}
    \caption{CAD cross Section highlighting the \SI{40}{\kelvin} stage (S1) thermal paths starting from the pulse tubes and meeting at the S1 flexible connection. The PT420-RM is on the left side of the diagram, and the PT425-RM is hidden behind the payload cryostat. The S1 components are highlighted in red.}
    \label{fig:40kpath}
\end{figure*}

The two pulse tubes independently provide cooling at the \SI{40}{\kelvin} stage (S1). The tens of watts of cooling power at this stage intercept as much heat as possible before it propagates to colder stages. The S1 thermal paths are highlighted in Figure~\ref{fig:overview} and Figure~\ref{fig:40kpath} in red. 

In the payload cryostat, the first stage of the PT425-RM is thermally connected to the S1 payload plate via C10100 oxygen-free high conductivity copper (C101 Cu) cold fingers and flexible thermal straps. The PT425-RM can be seen on the right side of Figure~\ref{fig:overview}. The S1 payload plate is an octagonal \SI{12}{\milli\meter}-thick C101 Cu plate supported by seven G10 tubes and enclosed by vertical C101 Cu flat panels that accommodate custom feedthroughs for wiring, the PT425-RM cold finger interface, and the DR interface. Wiring is routed through the S1 flat panels via MDM connectors on a PCB clamped on both sides with C101 Cu blocks to thermalize the lines. Flanges attached to the S1 flat panels mate to the cylindrical \SI{3}{\milli\meter}-thick Al6061 S1 radiation shield 3, which has an inner diameter of \SI{90}{\centi\meter} and a height of \SI{85}{\centi\meter}. Thirty layers of MLI protect S1 radiation shield 3 and the S1 payload plate from room temperature thermal radiation.

The first stage of the PT420-RM in the DR connects through a set of Cu braids to the S1 DR plate, which is mechanically supported by the vacuum chamber through four horizontal stainless steel (SS) tubes. A \SI{1}{\milli\meter}-thick Al radiation shield referred to as S1 radiation shield 1 extends approximately \SI{1}{\meter} from the S1 DR plate to the end of the shield. The end of the shield mates to a custom \SI{2}{\milli\meter}-thick Al6063 plate with a central aperture to allow the cold fingers to pass through, which in turn mates to the \SI{2}{\milli\meter}-thick Al6063 S1 radiation shield 2 with a length of \SI{0.3}{\meter}. The clearance between the compressed soft vacuum bellows and S1 radiation shield 2 is approximately \SI{6}{\milli\meter}. Together, these shields establish a continuous \SI{40}{\kelvin} stage radiation boundary that substantially reduces radiative loading at lower temperature stages.



\begin{figure*}[h!]
    \centering
    \includegraphics[width=1\textwidth]{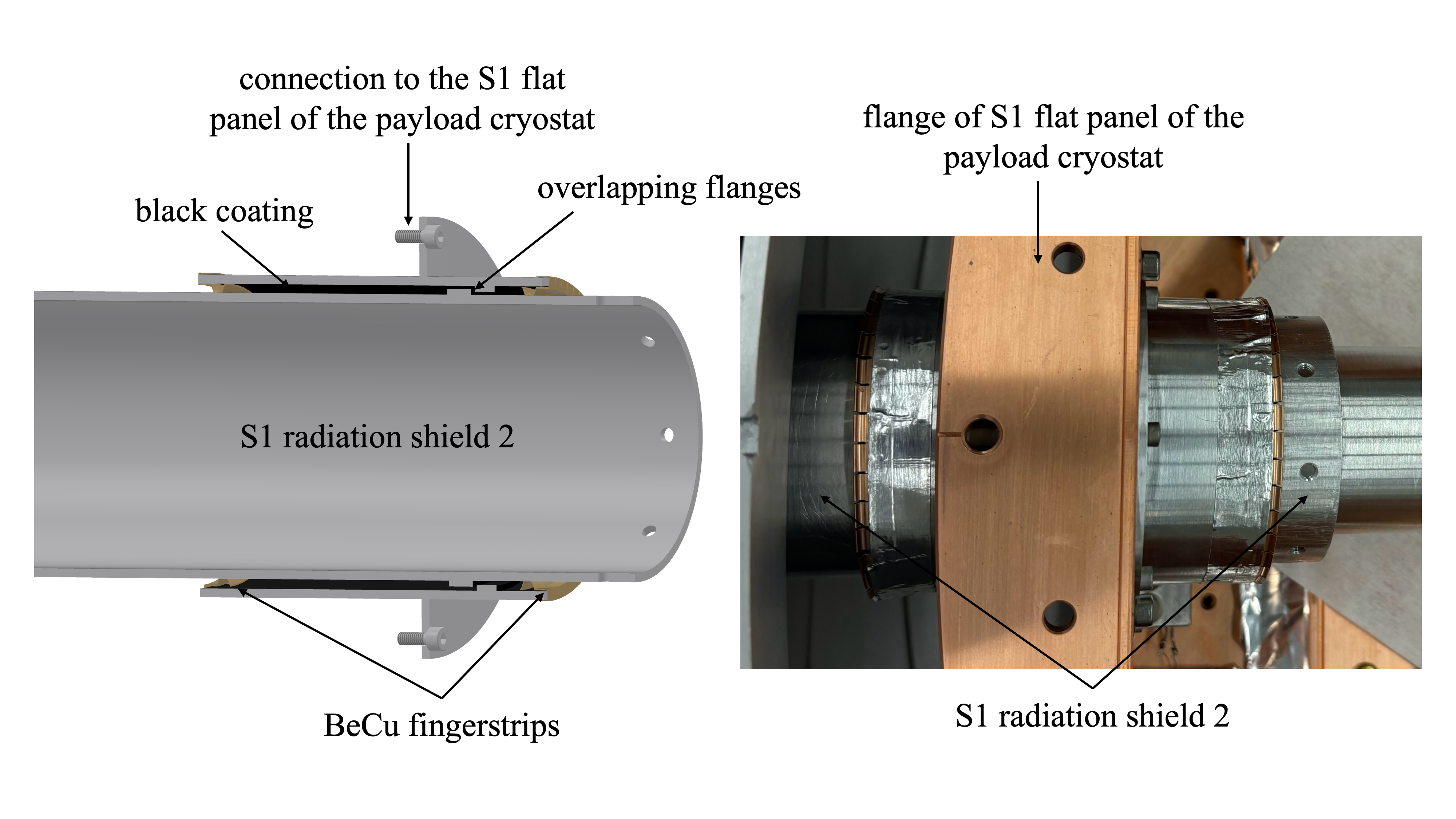}
    \caption{Light-blocking, mechanically-flexible connection between S1 radiation shield 2 of the DR and S1 flat panel of the payload cryostat. The inside surface of the tube and the outside surface of S1 radiation shield 2 are blackened and flanged to create a tortuous path that effectively prevents thermal radiation onto colder stages. The left image shows a labeled CAD rendering of a cross Section of the concentric cylinders and the right image is a photograph of the top view of the S1 flexible connection installed in the cryogenic system.}
    \label{fig:40kflex}
\end{figure*}
At this stage, the DR and payload cryostat thermal paths interface via the S1 flexible connection, labeled in Figure~\ref{fig:40kpath}. The flexible connection, shown in a CAD rendering and photograph in Figure~\ref{fig:40kflex}, provides radiation shielding and mechanical compliance. The outer cylinder of the flexible concentric connection is mounted on one of the S1 flat panels of the octagon. At both ends of the concentric cylinder, beryllium copper (BeCu) fingerstrips are employed. These strips compress during the insertion of S1 radiation shield 2, serving as a barrier to thermal radiation. The use of BeCu is specifically chosen for its mechanical flexibility at cryogenic temperatures, ensuring consistent contact even under motion. The inner surface of the outside cylinder is coated with a matte black highly absorptive material that attenuates stray thermal radiation~\cite{Boc1994}. Overlapping flanges on both the cylinder and the shield eliminate direct lines of sight from the warmer surfaces to the colder stages, providing an additional barrier to radiation.

\subsection{4 K stage (S2) thermal path \label{sec:secondstage}}
\begin{figure*}[h!]
    \centering
    \includegraphics[width=1\textwidth]{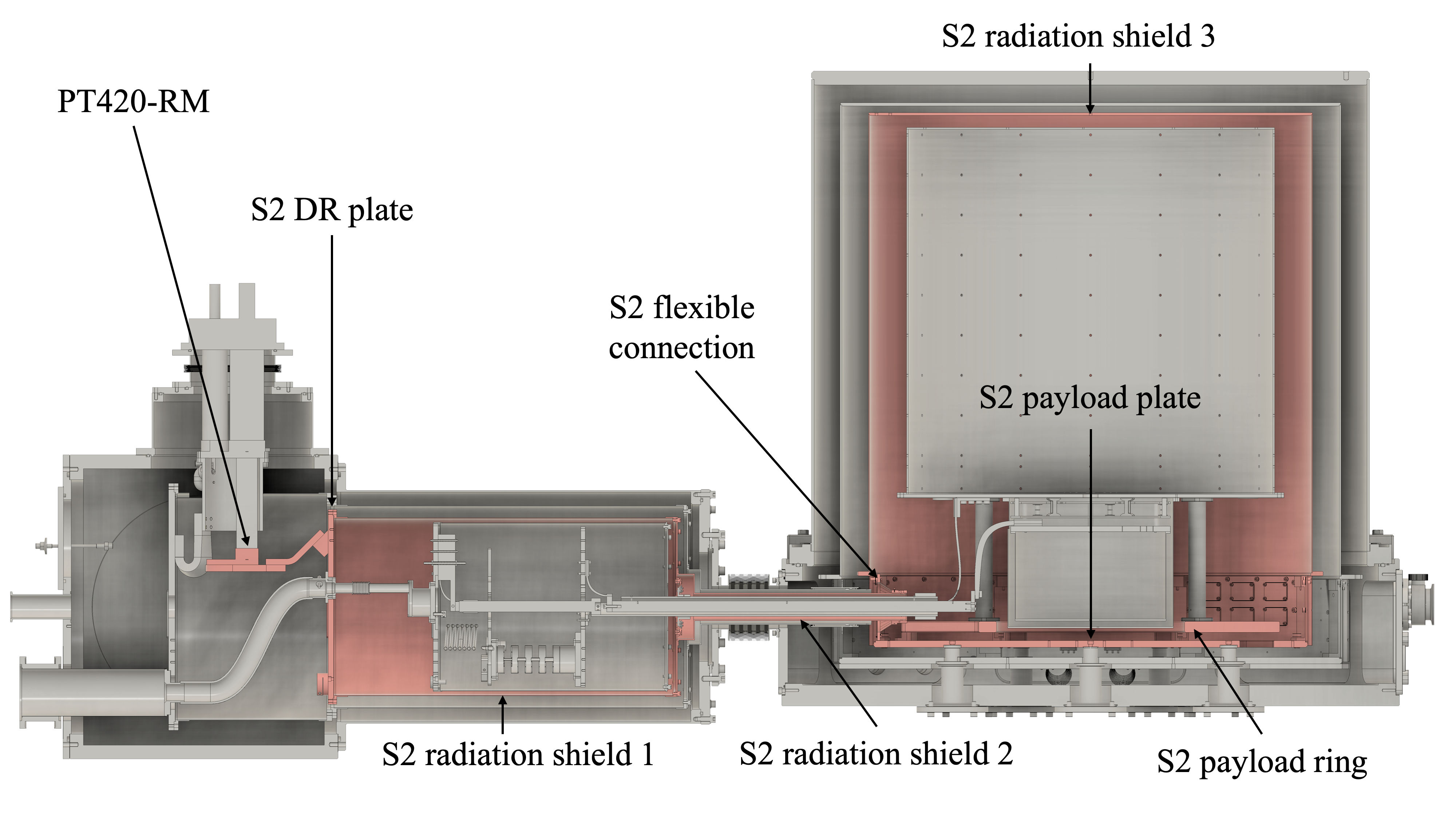}
    \caption{CAD cross Section highlighting the \SI{4}{\kelvin} stage (S2) thermal path starting from PT420-RM and PT425-RM and ending at S2 radiation shield 3 in the payload cryostat. PT420-RM is on the left side of the diagram, and PT425-RM is hidden behind the payload cryostat. The S2 components are highlighted in pink.}
    \label{fig:4kpath}
\end{figure*}

Both pulse tubes are thermally coupled at the \SI{4}{\kelvin} stage (S2) to cool components to approximately \SI{4}{\kelvin}. The S2 thermal path components are highlighted in Figure~\ref{fig:overview} and Figure~\ref{fig:4kpath} in pink. 

The PT425-RM second stage is connected to the S2 payload plate via C101 Cu cold fingers and thermal straps in the payload cryostat. The S2 payload plate is supported by the S1 payload plate through nested G10 and SS tubes at seven locations. As with S1, the octagonal \SI{12}{\milli\meter}-thick C101 Cu S2 payload plate is enclosed by eight vertical C101 Cu S2 flat panels that accommodate custom feedthroughs for wiring and cold fingers. A robust octagonal \SI{20}{\milli\meter}-thick Al7075 S2 payload ring is mounted using eight C101 Cu standoffs on the S2 payload plate. The S2 payload ring provides mechanical support for the lower temperature components and allows modular removal of the payload for improved access during payload assembly. The \SI{1.5}{\milli\meter}-thick C110 Cu S2 radiation shield 3 is attached to the flanges on the S2 flat panels and has an inner diameter and height of approximately \SI{80}{\centi\meter}. S2 radiation shield 3 is wrapped in reflective ultrahigh-vacuum foil with low emissivity to keep radiative heat loads on the shield low as the Cu oxidizes over time. 

In the DR, the second stage of the PT420-RM is connected to the S2 DR plate, as labeled in Figure~\ref{fig:4kpath}. The \SI{1}{\milli\meter}-thick S2 radiation shield 1 extends approximately \SI{60}{\centi\meter} from the S2 DR plate, ending with a custom \SI{2}{\milli\meter}-thick Al6063 plate. A narrower \SI{2}{\milli\meter}-thick Al6063 S2 radiation shield 2 extends the radiation shielding by approximately \SI{40}{\centi\meter} to the payload cryostat to shield the cold fingers as well as provide cooling power to the S2 components in the payload cryostat. The clearance between the S1 and S2 shields in the portion constrained by the vacuum bellows is only \SI{5}{\milli\meter}, so it is critical to maintain a high degree of concentricity. On the DR side, the shield is rigidly clamped to the bottom of S2 radiation shield 1. On the payload cryostat side, a rigid connection would transmit mechanical stress arising from differential thermal contraction. To avoid damage to the structures, the DR and payload cryostat \SI{4}{\kelvin} stage thermal paths interface via the S2 flexible connection. 

\begin{figure*}[h!]
    \centering
    \includegraphics[width=1\textwidth]{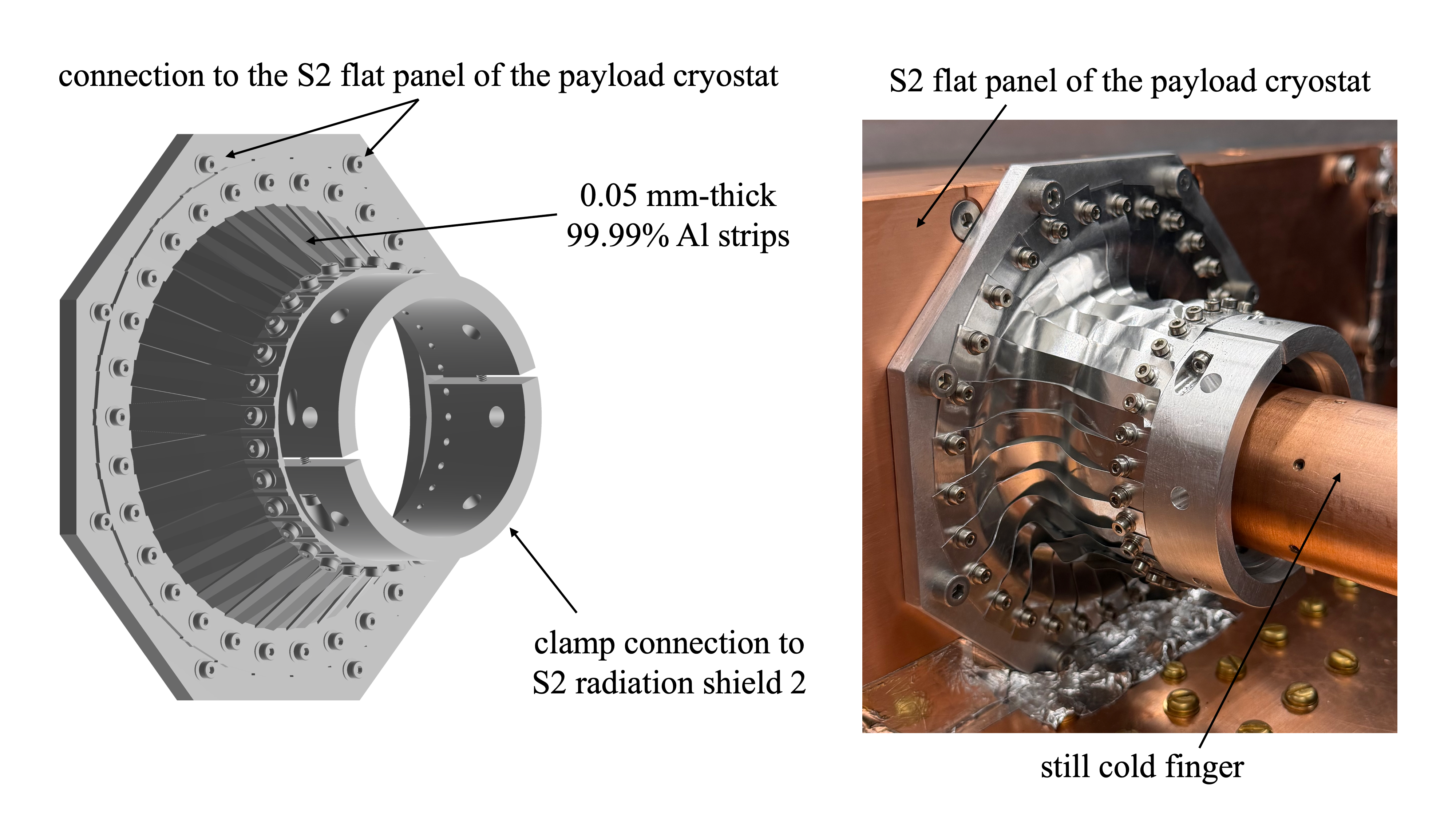}
    \caption{Thermally-conductive, mechanically-flexible connection between S2 radiation shield 2 and the S2 flat panel of the payload cryostat. The left image shows a labeled CAD rendering and the right image is a photograph of the flexible connection in situ.}
    \label{fig:4kflex}
\end{figure*}

The S2 flexible connection, shown in Figure~\ref{fig:4kflex}, is a  mechanically compliant and light-tight thermal link constructed from multiple overlapping \SI{0.05}{\milli\meter}-thick high-purity Al strips. These strips are arranged circumferentially to cover the full perimeter of the shield, and Al tape covers the inside of the strip arrangement. The cumulative cross-sectional area of the Al strips yields high enough thermal conductance for cooling power to be transmitted. Figure~\ref{fig:4kflex} shows a CAD rendering of the flexible connection assembly on the left and a photograph of the flexible connection integrated into the cryogenic system on the right. The outer panel of the flexible connection is mounted on one of the eight S2 flat panels, and the inner portion connects to S2 radiation shield 2 with a circular clamp.

The payload and cryogenic components that are eventually cooled to lower temperatures must be first pre-cooled to approximately \SI{4}{\kelvin} by this stage. During this initial cooling phase, the lower temperature components connected to the still and MXC plates are thermally linked to the S2 DR plate with heat switches built in to the SS support tubes in the DR. The heat switches become thermally non-conductive during the DR mixture condensing process. 

\subsection{Thermal path of the still \label{sec:still}}
\begin{figure*}[h!]
    \centering
    \includegraphics[width=1\textwidth]{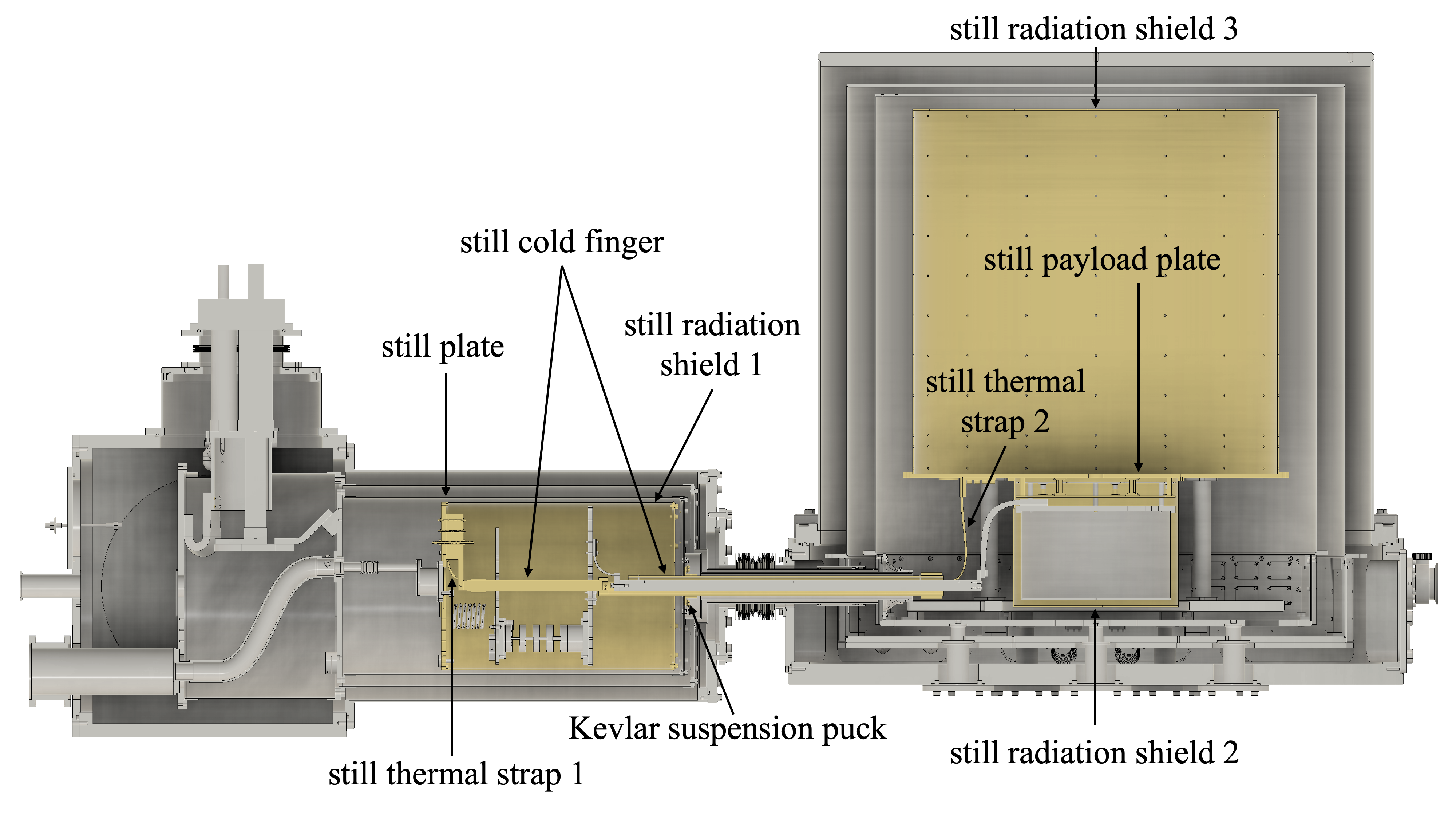}
    \caption{Still stage thermal path starting from the still plate in the DR on the left and ending at the still stage radiation shields in the vertical cryostat on the right. Still stage components are highlighted in gold.}
    \label{fig:1kpath}
\end{figure*}

The still stage is cooled by the Au-plated Cu still plate of the DR and is highlighted in Figure~\ref{fig:overview} and Figure~\ref{fig:1kpath} in gold.

In the payload cryostat, the still payload plate is supported by the S2 payload ring with four G10 tubes. The bulk of the still payload plate is constructed out of Al6061 and has a diameter of \SI{70}{\centi\meter} and thickness of approximately \SI{6}{\milli\meter}. Below the Al6061 plate, there is a \SI{1}{\milli\meter}-thick Au-plated C101 Cu plate of the same diameter to improve interface conduction of any connections and ensure minimal thermal gradient across the plate.

\begin{figure*}[h!]
    \centering
    \includegraphics[width=1\textwidth]{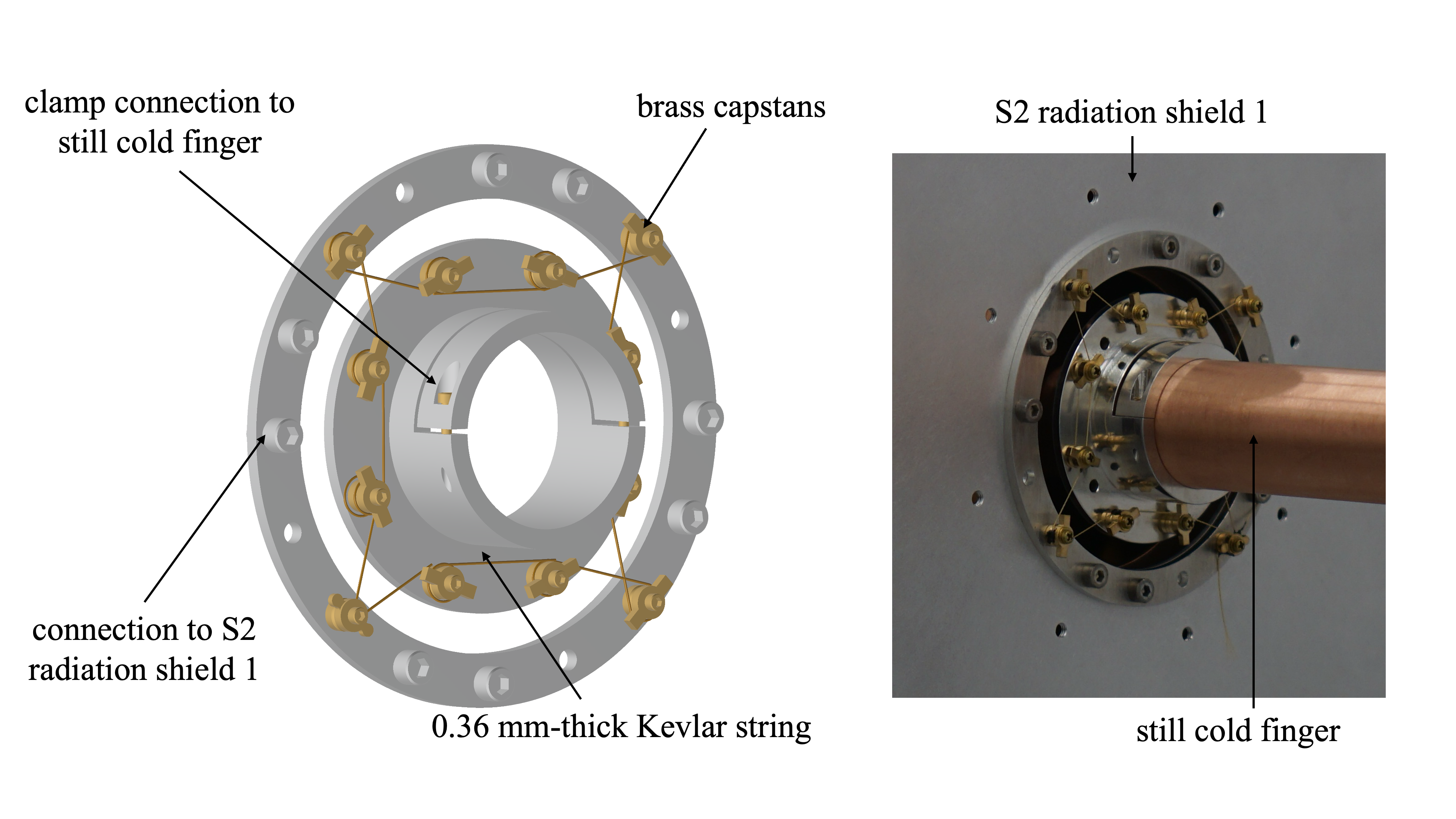}
    \caption{Kevlar suspension puck that provides thermally insulating mechanical support to the still cold finger. The left image is a CAD rendering and the right image is a photograph of the fully assembled puck in situ.}
    \label{fig:puck}
\end{figure*}

The still cold finger links the still plate in the DR to the still payload plate. The cold finger is divided into two C101 Cu sections joined by four M4 fasteners. The first section is a \SI{26}{cm}-long solid rod of diameter \SI{2}{cm} that is mechanically and thermally coupled to the DR still plate and is allowed to hinge. This hinged connection holds the still cold finger with a single M5 screw and allows the cold finger to rotate downward during installation, where it can rest on a temporary support fixture mounted on the MXC plate. The second section of the still cold finger consists of a \SI{63}{cm}-long tube with an inner diameter of \SI{3}{cm} and an outer diameter of \SI{3.6}{cm}. This geometry allows the base cold finger to be suspended concentrically inside the still cold finger, which maximizes the available space for cold fingers at both stages. Maintaining accurate alignment of the long horizontal cold fingers is essential to prevent contact between stages. An intermediate support of the still cold finger is implemented using a web of tensioned Kevlar string, forming a lightweight suspension structure referred to as a ``puck''~\cite{puckpatent}, shown in Figure~\ref{fig:puck}. The tensioned Kevlar string is wound around brass capstans positioned on two rings at different temperature stages. The inner ring of the puck clamps to the still cold finger, while the outer ring is connected to S2 radiation shield 1 at the \SI{4}{\kelvin} stage. Kevlar was selected for its exceptional tensile strength and extremely low thermal conductivity~\cite{Pob2007}. The tensioned-string design provides lateral stiffness sufficient to stabilize the cold finger while introducing negligible conductive heat load. 

\begin{figure*}[h]
    \centering
    \includegraphics[width=1\textwidth]{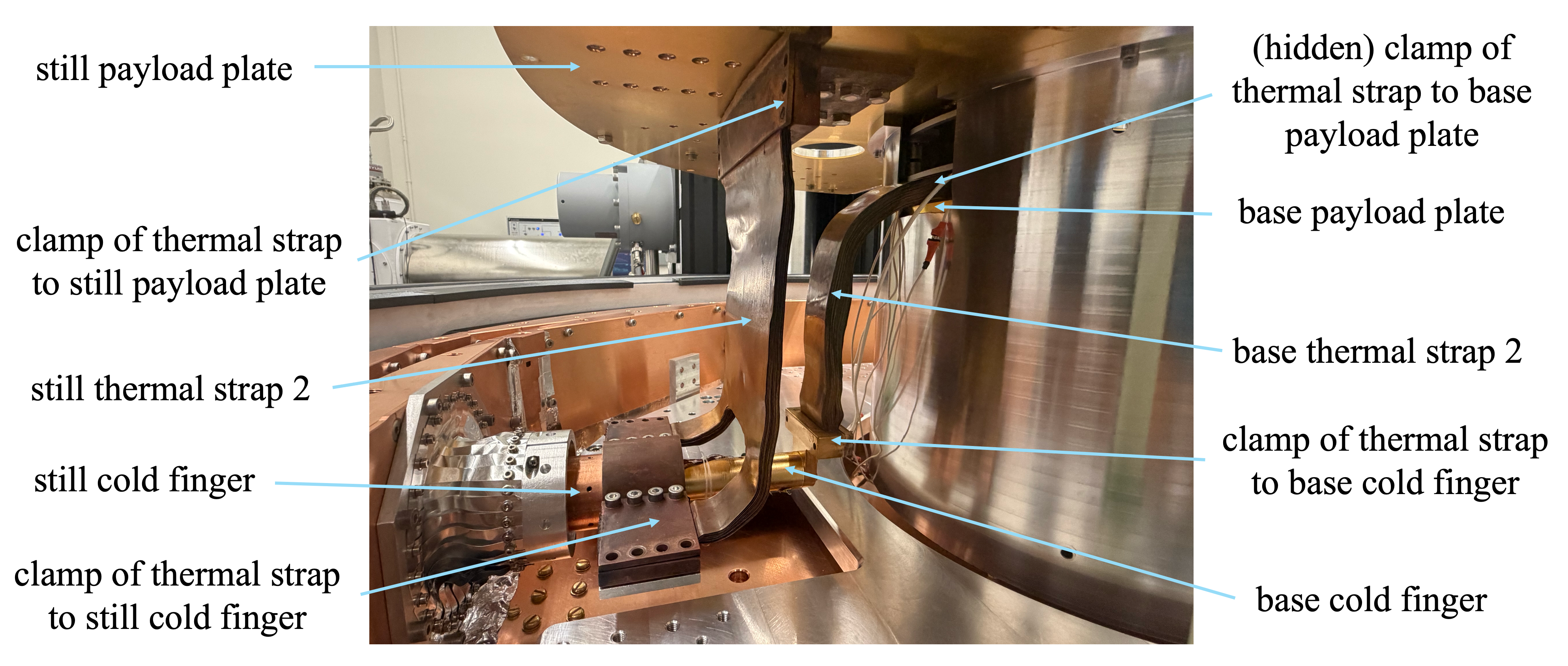}
    \caption{Photograph of the flexible thermal connections between the cold fingers on the left and the still payload plate and base payload plate at the top.}
    \label{fig:braids}
\end{figure*}

Thermal coupling on either side of the still cold finger is provided by still thermal strap 1 and still thermal strap 2, which are flexible Cu strap assemblies composed of sets of C110 Cu straps TIG welded to C101 Cu blocks. The strap geometry maximizes cross-sectional area for heat transport while remaining mechanically compliant. This flexibility is critical: it allows the cold finger to contract freely during cooldown. Both strap assemblies clamp on one side to a flat surface and on the other side to the circular still cold finger. Still thermal strap 2, visible in Figure~\ref{fig:braids}, has a cutout to accommodate its required curvature while preventing a touch between components of different temperatures.

Shielding at the still stage provides the final barrier against radiation and EMI reaching the base stage. In the DR, still radiation shield 1 surrounds the cold fingers up to the point where they enter the narrow region that passes through the bellows between the DR and payload cryostat. In the payload cryostat, the bottom of the still payload plate mates to still radiation shield 2, which is made of Al6061 and has a diameter of \SI{30}{\centi\meter} and a height of approximately \SI{24}{\centi\meter}. The inside of still radiation shield 2 is painted black to mitigate stray radiation. The Al1100 still radiation shield 3 mates to the top of the still payload plate and is \SI{66}{\centi\meter} in both diameter and height. 
\subsection{Thermal path of the base \label{sec:base}}


\begin{figure*}[h]
    \centering
    \includegraphics[width=1\textwidth]{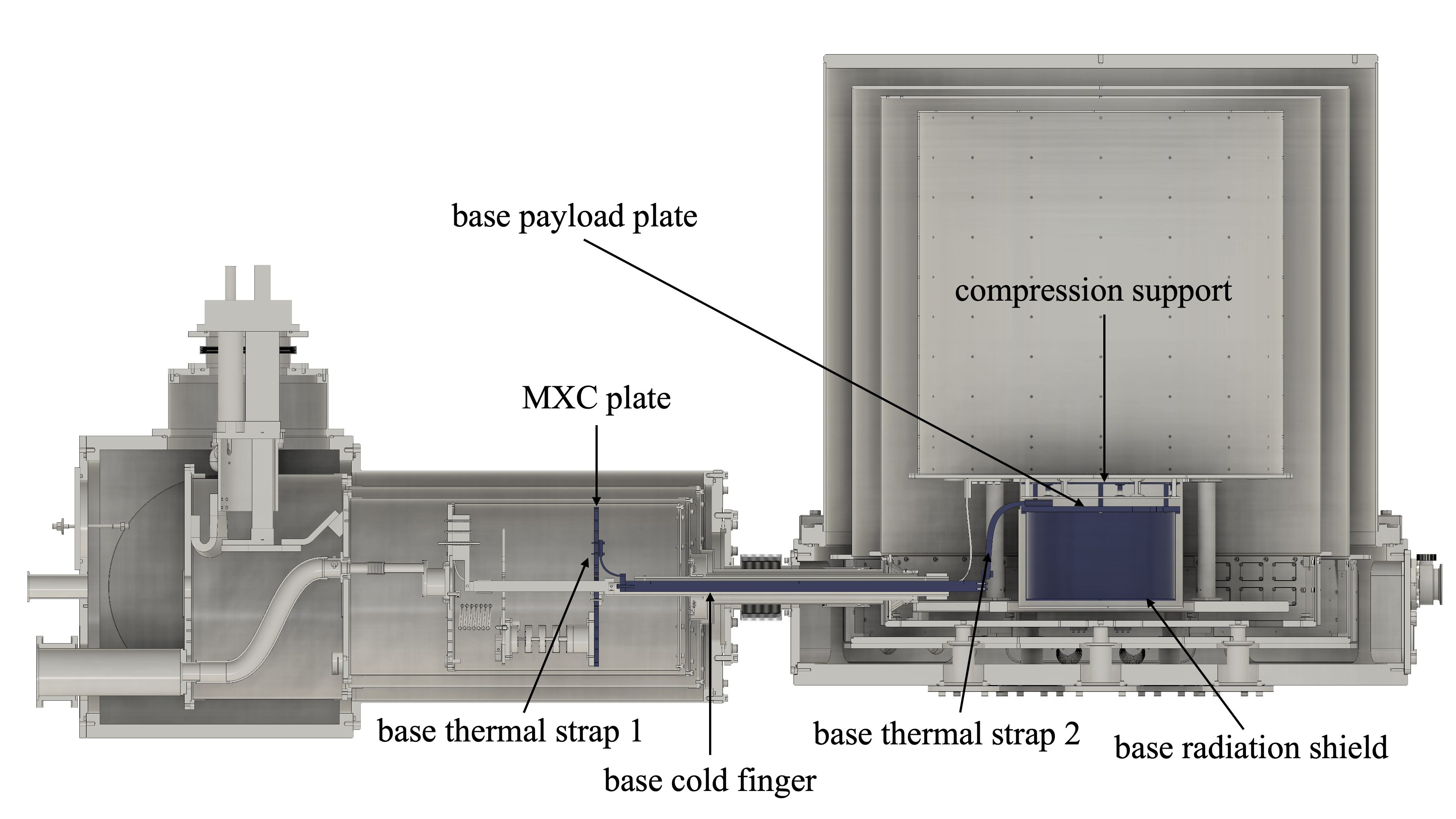}
    \caption{Base stage thermal path starting from the MXC plate in the DR on the left and ending at the base radiation shield in the payload cryostat on the right. Base stage components are highlighted in blue.}
    \label{fig:basepath}
\end{figure*}

\begin{figure*}[h]
    \centering
      \includegraphics[width=.9\textwidth]{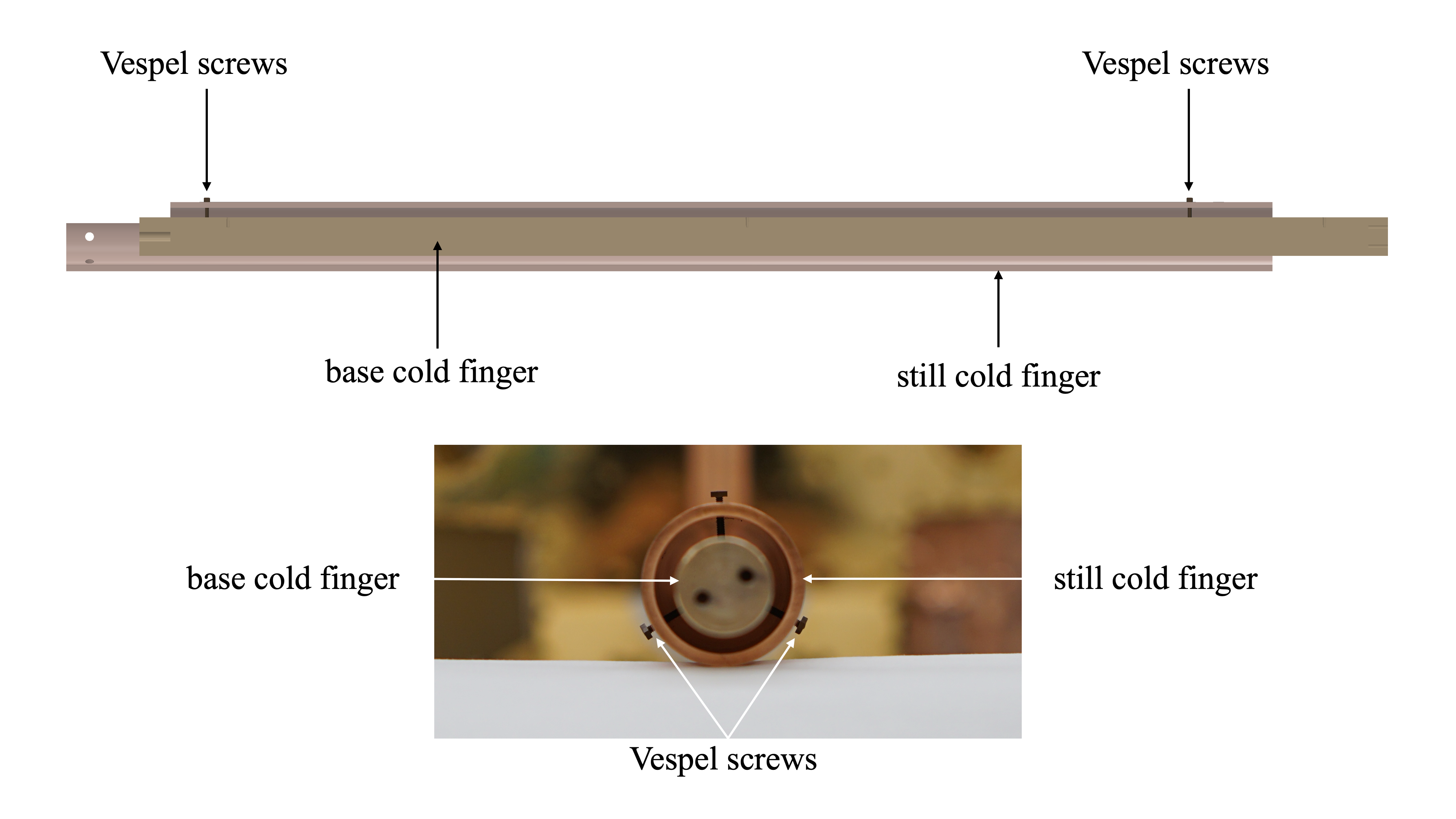}
    \caption{M2 Vespel SP-1 screws suspend the base cold finger inside the still cold finger tube. Top CAD render shows a cross section side view, pointing out the Vespel screw locations along the length of the base cold finger. The lower photograph shows Vespel screw positioning from the right. Only the lower two screws support the weight of the base cold finger; the top screw is an option to prevent a touch between temperature stages.}
    \label{fig:vespelscrews}
\end{figure*}

\begin{figure*}[h]
    \centering
    \includegraphics[width=1\textwidth]{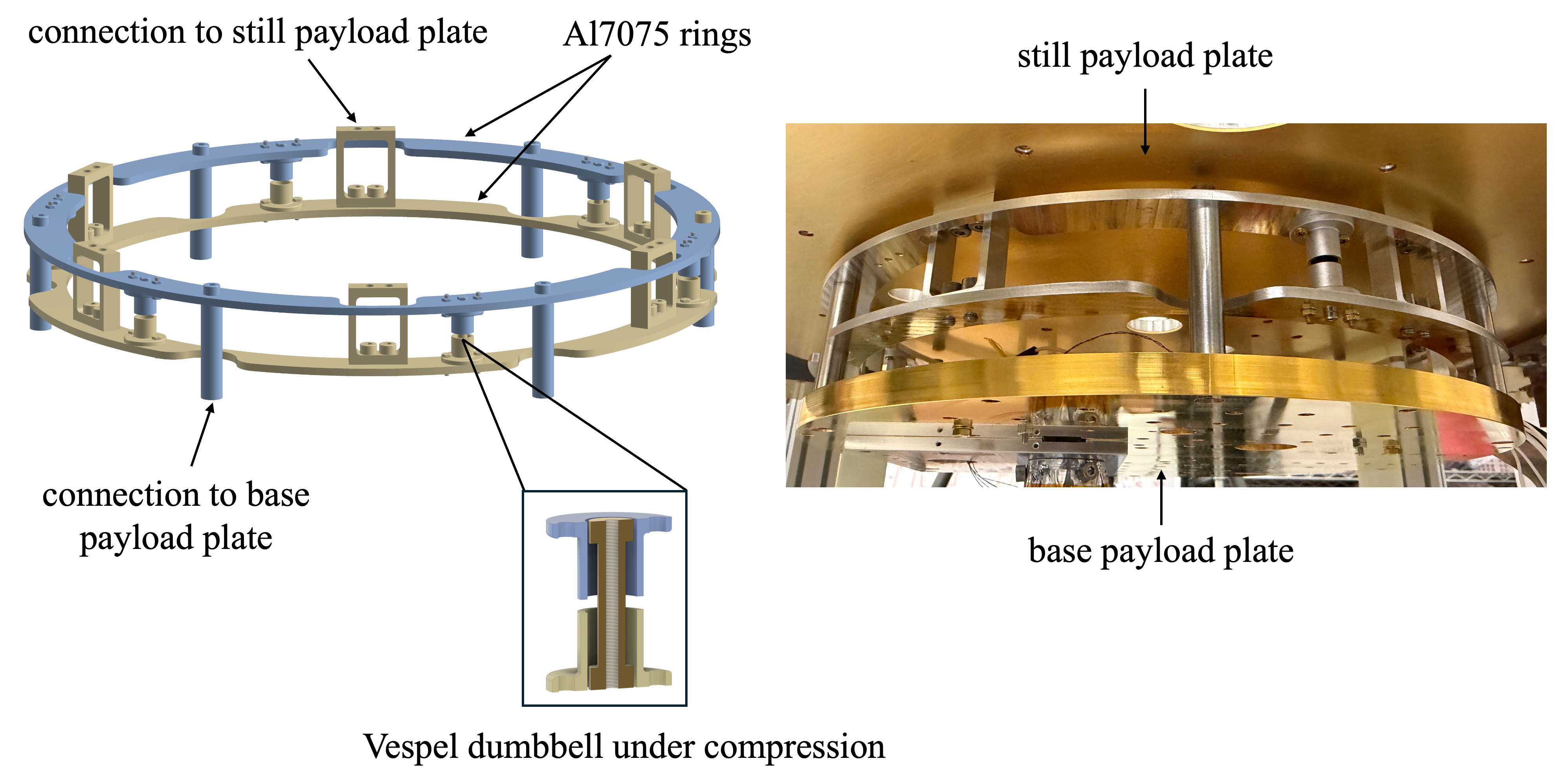}
    \caption{The compression support of the base payload plate with mechanically-supportive thermally-insulating Vespel SP-1 dumbbells under compressive load. The left image shows a CAD rendering of the design. The upper ring, highlighted in blue, is attached by six Al rods to the base payload plate and the lower ring, highlighted in gold, is attached by six box-shaped Al elements to the still payload plate. The Vespel dumbbells between the two rings act as standoffs between the base payload plate and still payload plate. The right image shows a photograph of the support connected to both the still and base payload plates.}
    \label{fig:compression}
\end{figure*}

The base payload plate in the payload cryostat interfaces with the resonator and its readout chain and therefore demands the lowest achievable temperature and highest thermal stability. Even small parasitic heat loads or temperature gradients can degrade the resonator quality factor or increase amplifier noise, making the thermal impedance between the MXC plate and the experimental payload a critical design constraint. The Au-plated C101 Cu \SI{10}{\milli\meter}-thick base payload plate is cooled by the MXC plate in the DR through a series of thermal components. The thermal path from the MXC plate of the DR to the base payload plate is highlighted in blue in Figures~\ref{fig:overview} and \ref{fig:basepath}. 

The base cold finger is the dominant thermal resistance between the MXC plate and the base payload plate. It is a \SI{65}{\centi\meter}-long, \SI{2}{\centi\meter}-thick Au-plated C101 Cu rod. Base thermal strap 1 connects via flat clamps to the end of the cold finger and the MXC plate. On the payload side, base thermal strap 2 connects the end of the cold finger to the top of the base payload plate. Base thermal strap 2 and its connection to the cold finger are shown in Figure~\ref{fig:braids}. Like the still thermal straps, the base thermal straps are made of C110 Cu strips TIG welded to C110 or C101 Cu flat clamps. All mating interfaces in the base thermal path are Au-plated to reduce interface resistance and ensure reproducible thermal performance. 

Mechanical support of the cold finger must minimize heat load while maintaining precise alignment. The rod is therefore suspended concentrically inside the still cold finger tube at two locations with two \SI{8}{mm}-long M2 screws fabricated from Vespel SP-1, an ultralow thermal conductivity polyimide with high mechanical strength. The mechanical support of the base cold finger within the still cold finger is shown in Figure~\ref{fig:vespelscrews}. The top screw acts as a spacer in the case of a misalignment and is not thermally connected to the base cold finger in normal operations. These supports are placed at both ends of the rod to maintain concentricity along the full length of the cold finger with minimal added heat load.

The base payload plate is mechanically supported and thermally isolated from the still plate using a compression-based Vespel SP-1 suspension depicted in Figure~\ref{fig:compression}. The thermally insulating, mechanically supportive elements are loaded in compression to take advantage of the high compressive strength of Vespel. Two interlocking support rings are fixed to the adjacent stages with Vespel tapped dumbbells compressed between them. In this configuration, the base stage detector mass is supported entirely by Vespel under compressive forces. This geometry additionally provides a passive fail-safe; in the event of Vespel support degradation, the upper ring can drop only \SI{1.8}{mm} before contacting the lower structure. The resulting design combines low thermal conductance, high mechanical stiffness, and robust failure tolerance, ensuring stable operation of the mK stage.

An Al6061 shield surrounds the space below the base payload plate to protect the sensitive electronics from radiation and EMI. The shield is \SI{16}{\centi\meter} tall and encloses the full volume beneath the base payload plate. 

The heat loads on the base payload plate are dominated by conduction through the mechanical supports. The Vespel dumbbell compression supports from the still payload plate are estimated to contribute \SI{0.75}{\micro\watt} per dumbbell, extrapolated from cryogenic conductivity data \cite{Moed2008}. Likewise, the Vespel screws supporting the base cold finger from the still cold finger are calculated to provide an additional \SI{0.66}{\micro\watt} of conductive heat per screw. The inductor centering support, a larger variation of the Kevlar suspension puck, is expected to provide an additional \SI{0.32}{\micro\watt} by conduction between the \SI{4}{\kelvin} magnet and base-temperature inductor \cite{50l_paper, Ventura2000}. The dominant radiative contribution from the \SI{4}{\kelvin} magnet to the inductor is calculated to be \SI{0.7}{\micro\watt}; this load is suppressed by the low emissivity of Al and Nb at cryogenic temperatures \cite{saini2015estimation, baudouy2015heat, 50l_paper}. 

Wiring is thermalized and fed through the base payload plate to the shielded environment similarly to the higher temperature stages. Thermalization PCBs are clamped with Cu blocks to the plate and fed through with MDMs or smaller connectors for individual twisted pairs. 


\subsection{Cryostat controls and readout}
\begin{figure*}[h!]
    \centering
    \includegraphics[width=1\textwidth]{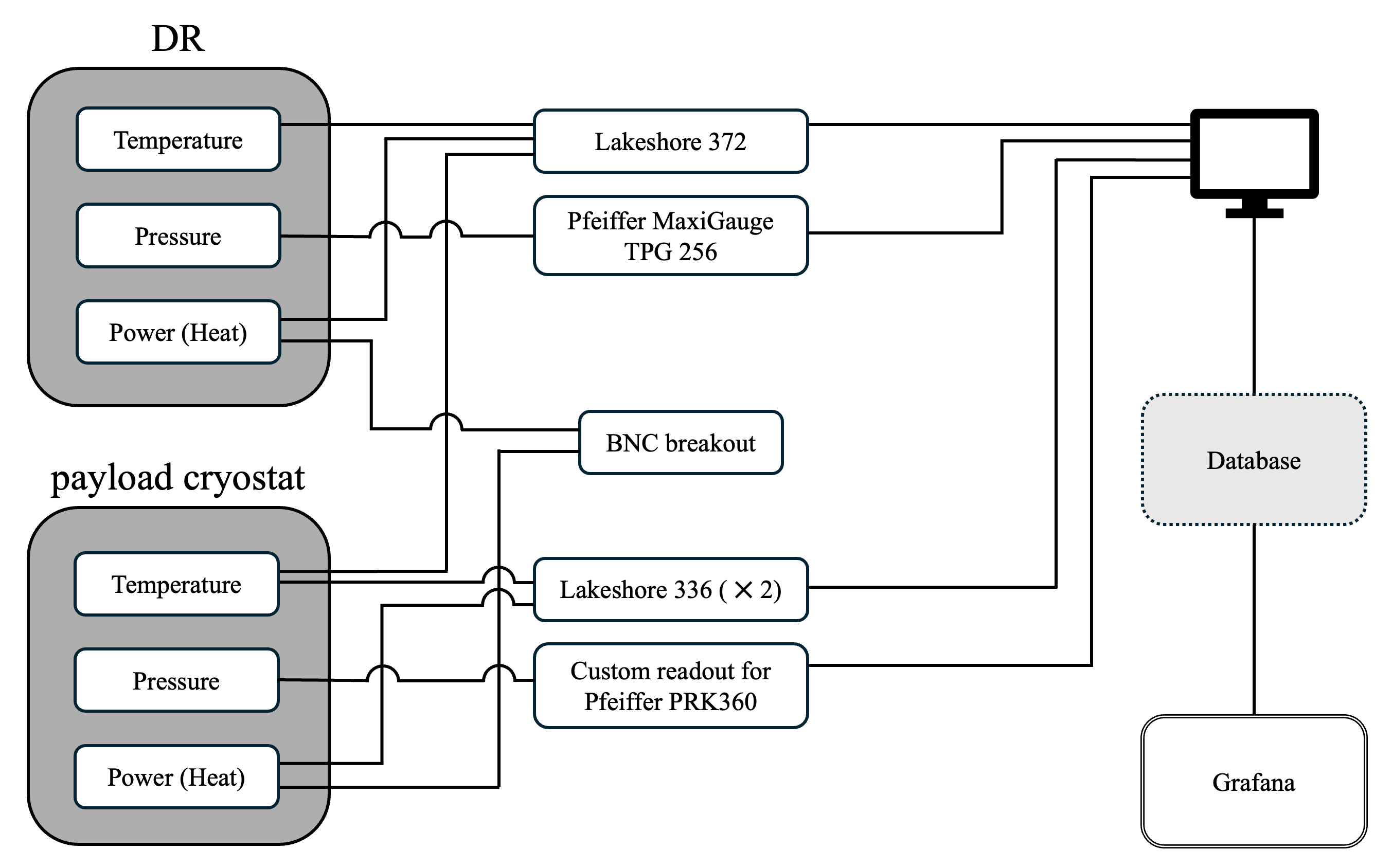}
    \caption{Overview of the slow monitoring and control readout architecture for the cryogenic system. The DR and payload cryostat have independent temperature and pressure sensors as well as inputs for heater power. Temperature and pressure in the DR are read out by a Lakeshore 372 AC resistance bridge and a Pfeiffer MaxiGauge TPG 256 multi-gauge controller, respectively. A BNC breakout panel provides additional heater power inputs at room temperature. Temperature and heater power in the payload cryostat are handled by two Lakeshore 336 temperature controllers in addition to the Lakeshore 372, while its pressure is monitored by a Pfeiffer PRK360 gauge. All instruments communicate with a central computer, which logs data to a database. Recorded data is displayed and monitored in real time via a Grafana dashboard.}
    \label{fig:slowcontrols}
\end{figure*}

Reliable operation and characterization of the cryogenic system requires continuous monitoring and control of the temperatures, pressures, and applied heats at all stages of the system. In addition to verifying safe operating conditions, this instrumentation is essential for quantifying thermal loads, measuring specific conductances along the thermal paths and diagnosing unexpected heat loads during commissioning and steady-state operation. An overview of the readout system is shown in Figure \ref{fig:slowcontrols}.

Temperature sensors are distributed throughout the DR and the payload cryostat to provide complete coverage from room temperature down to the MXC. Sensor types are selected and calibrated according to the expected operating range to optimize sensitivity and accuracy. A complete list of sensors used is included in Table~\ref{table:base_temps}. Sensors interface with a LakeShore Cryotronics Model 372 resistance bridge or one of two LakeShore Cryotronics Model 336 controllers. Sensor wiring below 4 K is carefully thermalized at intermediate plates and routed using superconducting CuNi-clad NbTi twisted pairs to limit parasitic conductive loading. 

Resistive heaters are mounted at key locations throughout the thermal paths, including on the pulse tube stages, along the still stage, and along the base stage. In places where mounting with screws is possible, wirewound chassis mount resistors are used. Otherwise, surface mount resistors are soldered onto small sheets of Cu which are varnished onto the desired locations. These heaters enable controlled heat injection for active temperature stabilization as well as quantitative measurements of thermal conductance. By applying known heater power and recording the corresponding temperatures, we directly determine the effective thermal impedance of solid segments such as the cold fingers, flexible components such as straps, and interfaces; such measurements are presented in Section~\ref{sec:performance}. Heater leads are routed to room temperature and thermalized using the same practices as sensor wiring to prevent unintended heat leaks. Power is supplied through battery-powered voltage sources for ground loop mitigation.

Vacuum and DR pressures are continuously monitored to ensure proper cryostat performance. These signals are collected using a Pfeiffer Vacuum MaxiGauge TPG 256 controller, providing coverage from rough vacuum through high-vacuum regimes. 

All temperature, pressure, and heater power channels are integrated into a centralized slow-controls system. Analog and digital signals are routed through a breakout interface and recorded to a PostgreSQL database for long-term storage. Live monitoring and visualization are provided through the platform Grafana \cite{Grafana}, which enables real-time dashboards, automated alerts, and trend analysis. This infrastructure supports both operational oversight and offline analysis. The resulting dataset is also used to validate thermal models and to guide iterative improvements to mechanical interfaces and shielding. This readout architecture provides precise, low-noise monitoring of all relevant thermodynamic variables while maintaining minimal parasitic heat loads, ensuring both reliable operation and accurate thermal characterization of the cryogenic system.

\subsection{Design for earthquake survivability}
The system is located at Stanford University, which lies in a seismically active region of California, and earthquake survivability was therefore treated as an explicit engineering requirement. The guiding design philosophy prioritized conservative structural margins with the expectation that a seismic event could subject the apparatus to forces well beyond typical operating conditions. Both the DR and the payload cryostat are seismically bolted to the ground, preventing displacement or tipping during ground motion. The Vespel compression support of the base payload plate was load tested cryogenically using liquid nitrogen to verify mechanical integrity at low temperatures. Two additional steel supports are installed within the DR between the S2 DR plate and still plate to reinforce its internal structure against lateral and vertical loading.

The robustness of this design was validated on April 2, 2026, when a magnitude 4.6 earthquake struck near Brookdale, CA \cite{usgs_brookdale_2026}. The system sustained no structural damage; however, a transient disruption of the base temperature was observed, accompanied by fluctuations in temperature and pressure. Upon subsequent disassembly, the BeCu fingerstrips were found to be distorted. This suggests that the observed fluctuations were caused by mechanical relaxation of internal components following earthquake-induced movement. The fingerstrips were readjusted and the system returned to normal operation in the next cooldown. 
\section{Performance}
\label{sec:performance}

The cryogenic system was commissioned to verify that the system meets the cooling power and temperature requirements summarized in Section~\ref{sec:reqs}. Performance was evaluated in two parts: (i) transient cooldown measurements to characterize cooling time and thermal coupling between the DR and the payload cryostat, described in Section \ref{sec:transient}, and (ii) steady-state operation to quantify base temperature, cooling power, and agreement with the thermal model, described in Section~\ref{sec:steady-state}. Overall, the performance of the cryogenic system proved to satisfy the thermal design requirements and demonstrated preparedness for successful DMRadio-50L operation.

\subsection{Cooldown Characterization\label{sec:transient}}

\begin{figure}[H]
    \centering
    \includegraphics[width=.95\textwidth]{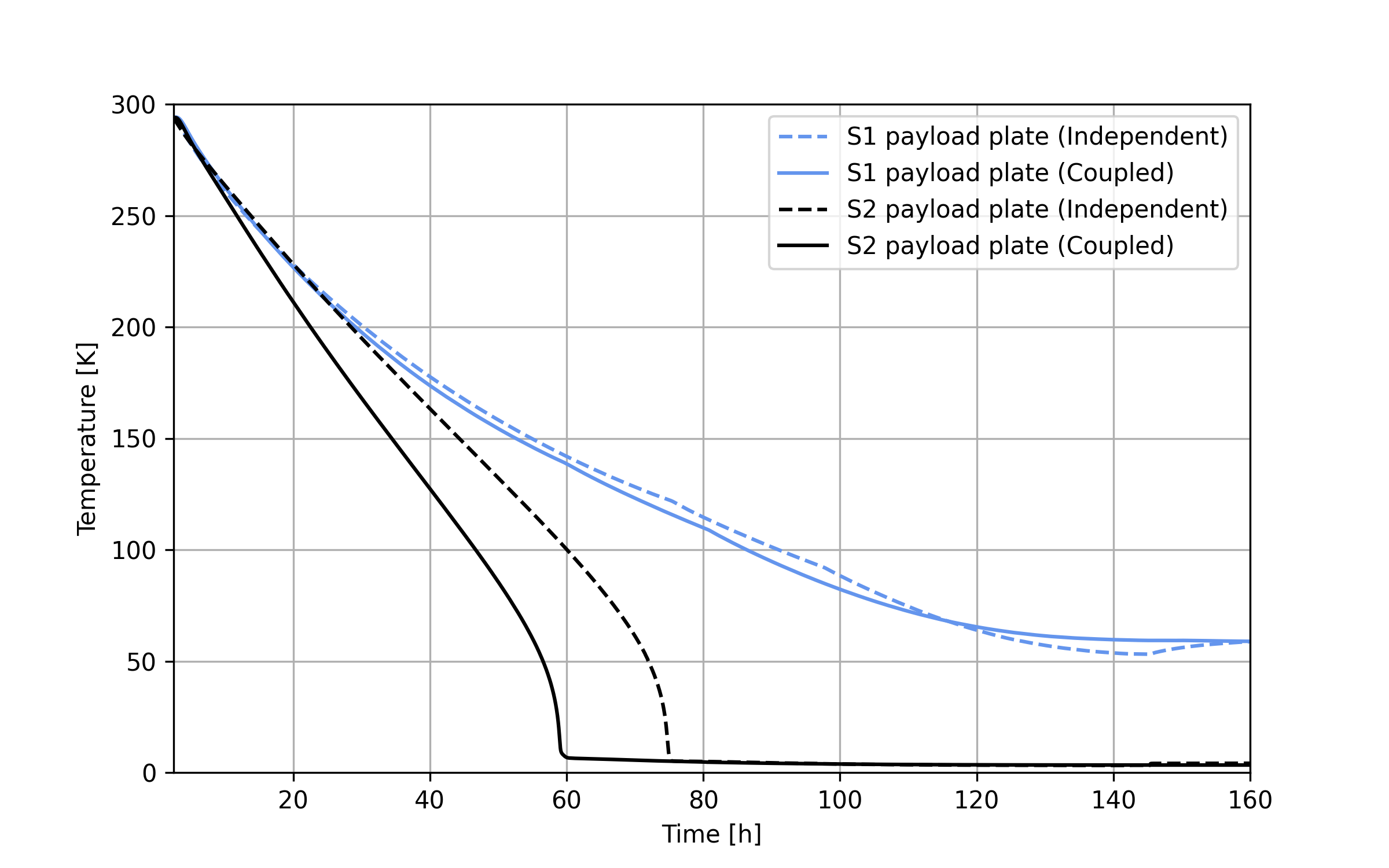} 
    \caption{Temperature of the S1 and S2 payload plates as a function of time, comparing cooling times during independent operation of the payload cryostat (Independent) to cooling times in the connected system (Coupled). The significant speed up in the cooling time of the S2 payload plate at the \SI{4}{\kelvin} stage is attributed to the contribution of the cooling power from the DR.}
    \label{fig:cooling_plot}
\end{figure}

Cooldown performance was measured from room temperature following vacuum pumping of all vacuum spaces. Figure~\ref{fig:cooling_plot} compares cooling trajectories for two configurations: independent operation of the payload cryostat and coupled operation of the payload cryostat with the DR. With only the payload cryostat active, the S1 payload plate and the S2 payload plate reach their steady-state temperatures within approximately \SI{150}{\hour} and \SI{75}{\hour}, respectively. When the DR is thermally coupled, the additional cooling power on the \SI{4}{\kelvin} stage reduces the cooling time from approximately \SI{75}{\hour} to \SI{60}{\hour}. Since we expect that the cooling bottleneck of the system with the full payload will be the \SI{4}{\kelvin} stage, the significantly greater cooling time of the \SI{40}{\kelvin} stage in its current configuration is not of concern.
\subsection{Steady-State Thermal Performance\label{sec:steady-state}}
\begin{figure}[h]
    \centering
    \includegraphics[width=1\textwidth]{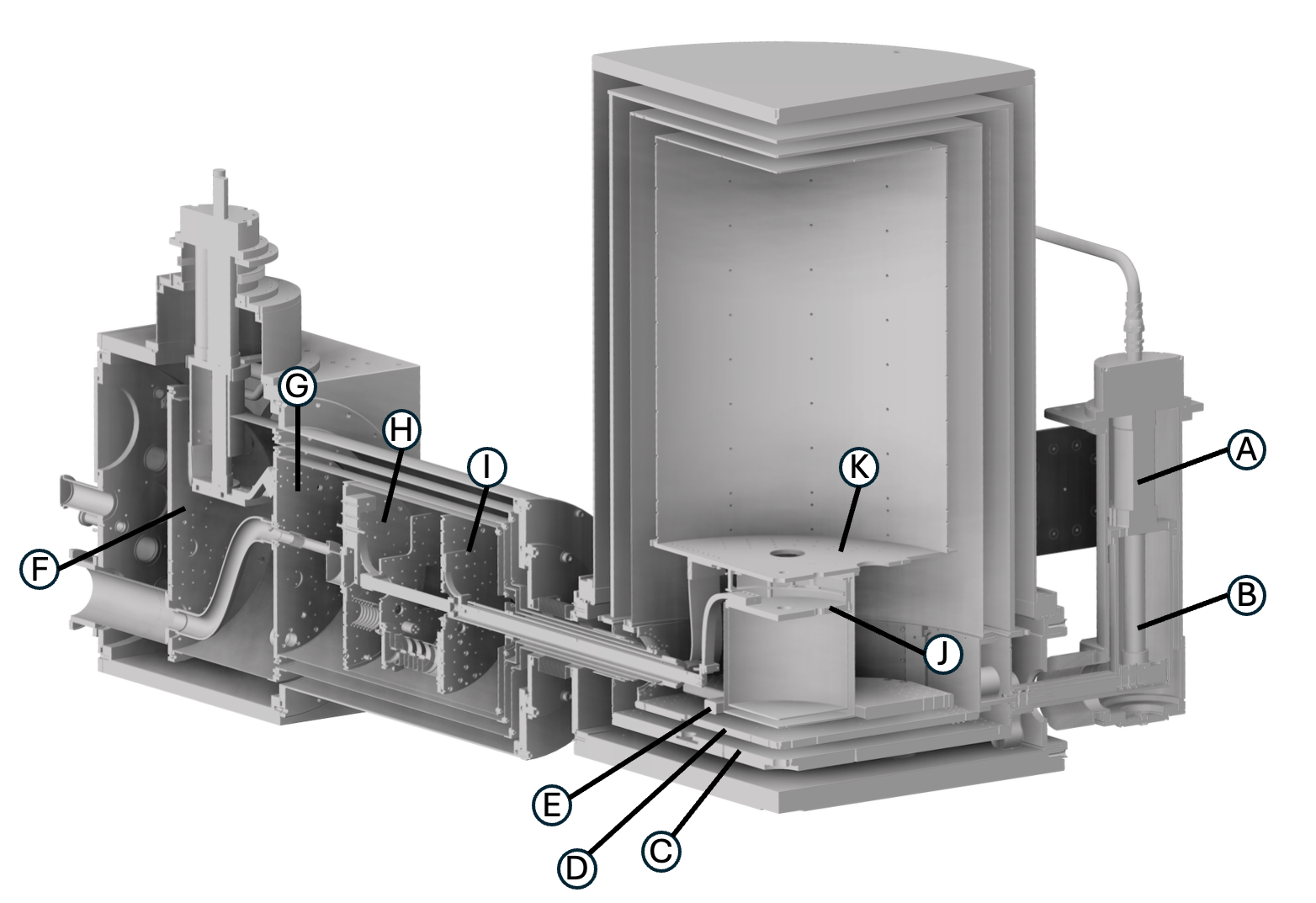}
    \caption{Temperature labels identified in Table~\ref{table:base_temps}.}
    \label{fig:labeled_overview}
\end{figure}
\begin{table}[h]
\centering
\caption{Measured temperatures at the \SI{40}{\kelvin} (S1), \SI{4}{\kelvin} (S2), still, and base stages of the cryogenic system.}
\begin{tabular}{lllll}
\toprule
Label & Cryostat & Location / Component & Temperature & Thermometer\\
\midrule
A & Payload cryostat & PT425-RM S1 & \SI{31.2}{\kelvin} & Si diode\\
B & Payload cryostat & PT425-RM S2 & \SI{2.4}{\kelvin} & Si diode\\
C & Payload cryostat & S1 payload plate & \SI{52.0}{\kelvin} & Cernox\\
D & Payload cryostat & S2 payload plate & \SI{3.4}{\kelvin} & Cernox\\
E & Payload cryostat & S2 payload ring & \SI{3.4}{\kelvin} & Cernox\\
F & DR & S1 DR plate & \SI{42.2}{\kelvin} & Platinum sensor\\
G & DR & S2 DR plate & \SI{2.7}{\kelvin} & Cernox\\
H & DR & still plate & \SI{571}{\milli\kelvin} & Cernox\\
I & DR & MXC plate & \SI{21}{\milli\kelvin} & RuO$_2$\\
\midrule
J & Payload cryostat & base payload plate & \SI{36}{\milli\kelvin} $\pm$ \SI{5}{\milli\kelvin} &  RuO$_2$\\
K & Payload cryostat & still payload plate & \SI{696}{\milli\kelvin} $\pm$ \SI{5}{\milli\kelvin} & RuO$_2$\\
\bottomrule
\end{tabular}
\label{table:base_temps}
\end{table}

After precooling with the pulse tubes, the cryogenic system reaches steady-state temperatures with the DR mixture in circulation and both pulse tubes running. Steady-state performance was evaluated by monitoring temperatures and applied heater powers to characterize thermal paths, quantify static heat loads, and compare measured temperatures to the design thermal model.

All stages reached stable operating temperatures well below the experimental requirements. The first and second stages of the pulse tube cryocooler PT425-RM stabilized at \SI{31.2}{\kelvin} and \SI{2.4}{\kelvin}, respectively, and the S2 payload plate and ring both settled around  \SI{3.4}{\kelvin}. The DR still plate operated at \SI{571}{\milli\kelvin}, and the MXC plate reached a base temperature of \SI{21}{\milli\kelvin}. The still and base payload plates stabilized at \SI{696}{\milli\kelvin} and \SI{36}{\milli\kelvin}, respectively. These temperatures are summarized in Table~\ref{table:base_temps}, and corresponding temperature sensor locations are shown in Figure~\ref{fig:labeled_overview}.

The measured base temperature of the MXC is consistent with the expected heat loads from wiring, supports, and radiation. While currently acceptable, the level of radiation can be improved for even lower base temperatures (see Section~\ref{sec:conclusion}). The base payload plate remains below \SI{50}{\milli\kelvin}, confirming that the cold-finger and strap assemblies provide sufficient thermal conductance to maintain a small enough temperature gradient between the MXC and the payload volume.

\begin{figure}[h]
    \centering
    \begin{subfigure}[b]{0.49\textwidth}
        \centering
        \includegraphics[width=\textwidth]{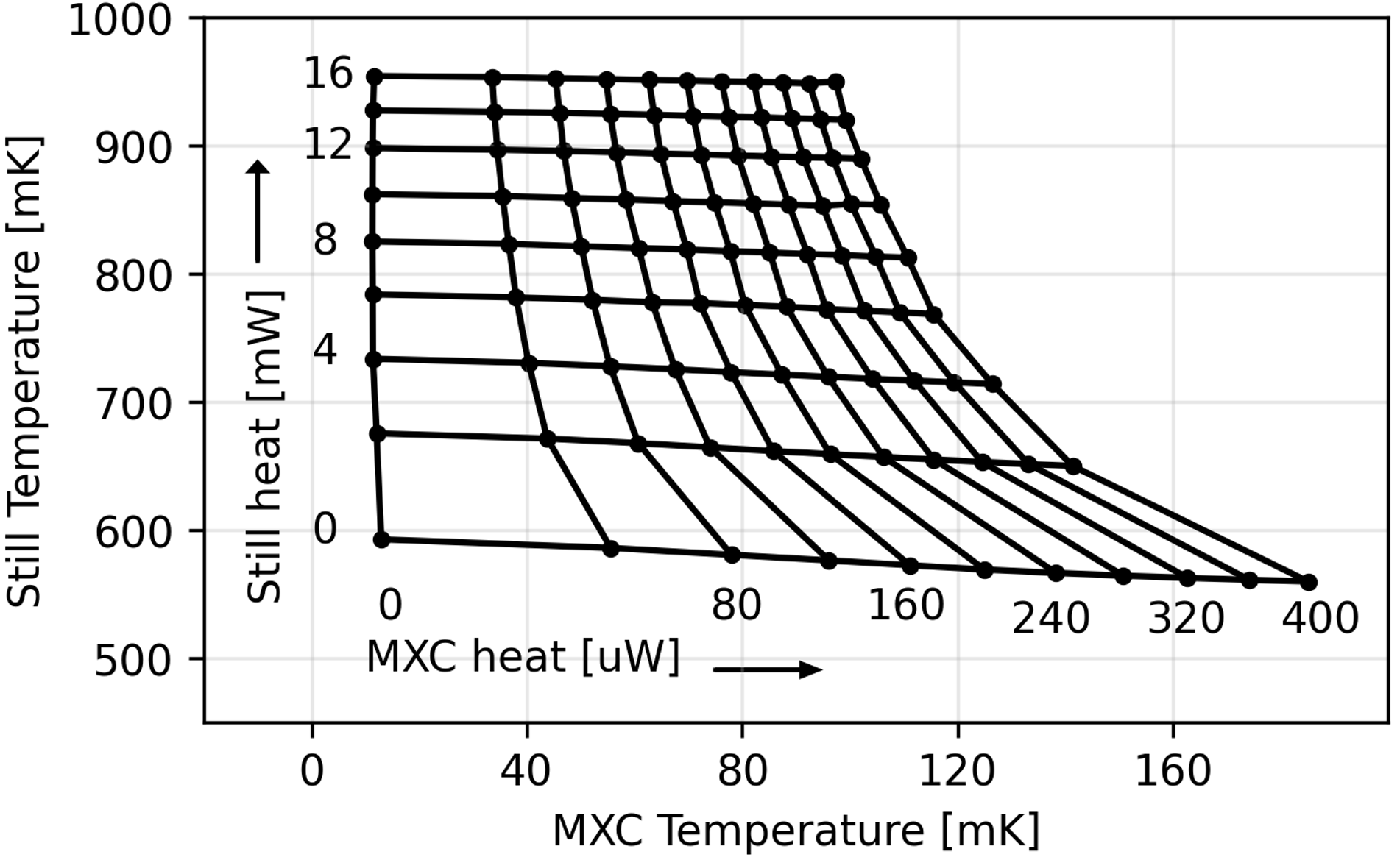}
        \caption{}
        \label{fig:DR_heat_map_a}
    \end{subfigure}
    \hfill
    \begin{subfigure}[b]{0.49\textwidth}
        \centering
        \includegraphics[width=\textwidth]{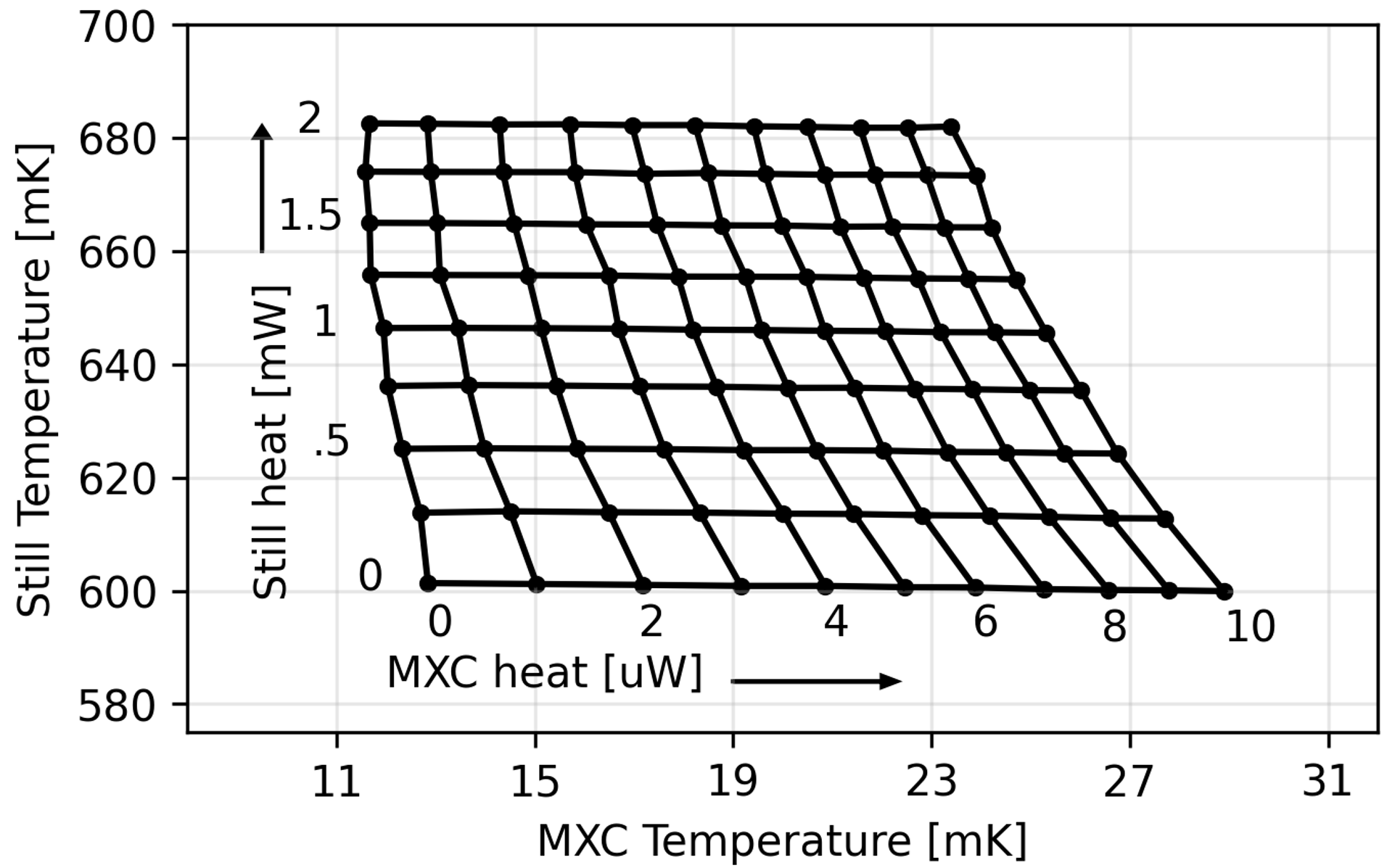}
        \caption{}
        \label{fig:DR_heat_map_b}
    \end{subfigure}
    \caption{Temperatures of the MXC plate and still plate in the DR at various applied heat loads with the cold fingers attached to the MXC plate and still plate. The axes represent the plate temperatures and the numbers in the plots represent the applied heat to either the base or still stages.}
    \label{fig:DR_heat_map}
\end{figure}

\begin{figure*}[ht]
    \centering
    \includegraphics[width=0.55\textwidth]{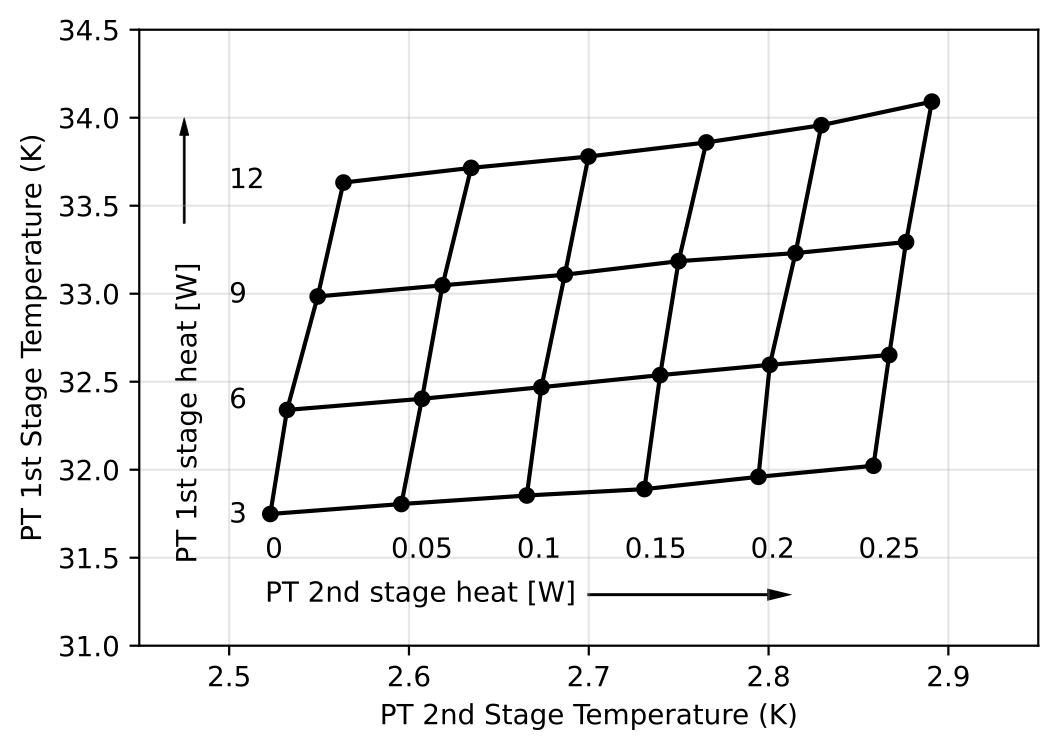} 
    \caption{Temperatures of the integrated \SI{40}{\kelvin} stage (1st stage) and \SI{4}{\kelvin} stage (2nd stage) of the PT425-RM in the payload cryostat at various applied heat loads.
    }
    \label{fig:PT_heat_map}
\end{figure*}

The cooling power available at each stage depends not only on its own temperature but also on the temperatures of adjacent stages. The \SI{40}{\kelvin} and \SI{4}{\kelvin}  stages are thermally coupled to each other, as are the still and base stages, meaning the thermal load on one stage directly affects the performance of the other. To quantify the relationship between cooling power and temperature, heat load maps were created, shown in Figure~\ref{fig:DR_heat_map} and Figure~\ref{fig:PT_heat_map}. A grid of operating points was constructed by systematically varying the heater power applied to two stages at a time, yielding corresponding temperature and heat load measurements at each stage. The result allows us to estimate the heat loads from the relative temperatures of the stages as more detector components are incorporated. Figure~\ref{fig:DR_heat_map} shows the thermal behavior of the still and MXC plates in the DR. Figure~\ref{fig:DR_heat_map_a} shows the MXC plate  exposed to applied heater power of 0-\SI{400}{\micro\watt} in increments of \SI{40}{\micro\watt} and the still plate exposed to applied heater power of 0-\SI{16}{\milli\watt} in increments of \SI{2}{\milli\watt}. Figure~\ref{fig:DR_heat_map_b} shows much smaller increments of applied heat to the still and MXC plates close to the range expected with the payload. For predicted heat loads at each stage, see Table~\ref{tab:summary}. The DR heat map was measured independently of the payload cryostat with the cold fingers connected to both the base and still stages to achieve a result as close as possible to the final expected heat load configuration while retaining a baseline for radiation and other unexpected heat loads. Figure~\ref{fig:PT_heat_map} is a heat load map of the \SI{40}{\kelvin} and \SI{4}{\kelvin} stages in a configuration where both cryogenic subsystems are connected and the still payload plate and still radiation shield 2 are installed and connected to the still cold finger via still thermal strap 2. 

\begin{figure}[h]
    \centering
    \includegraphics[width=.9\textwidth]{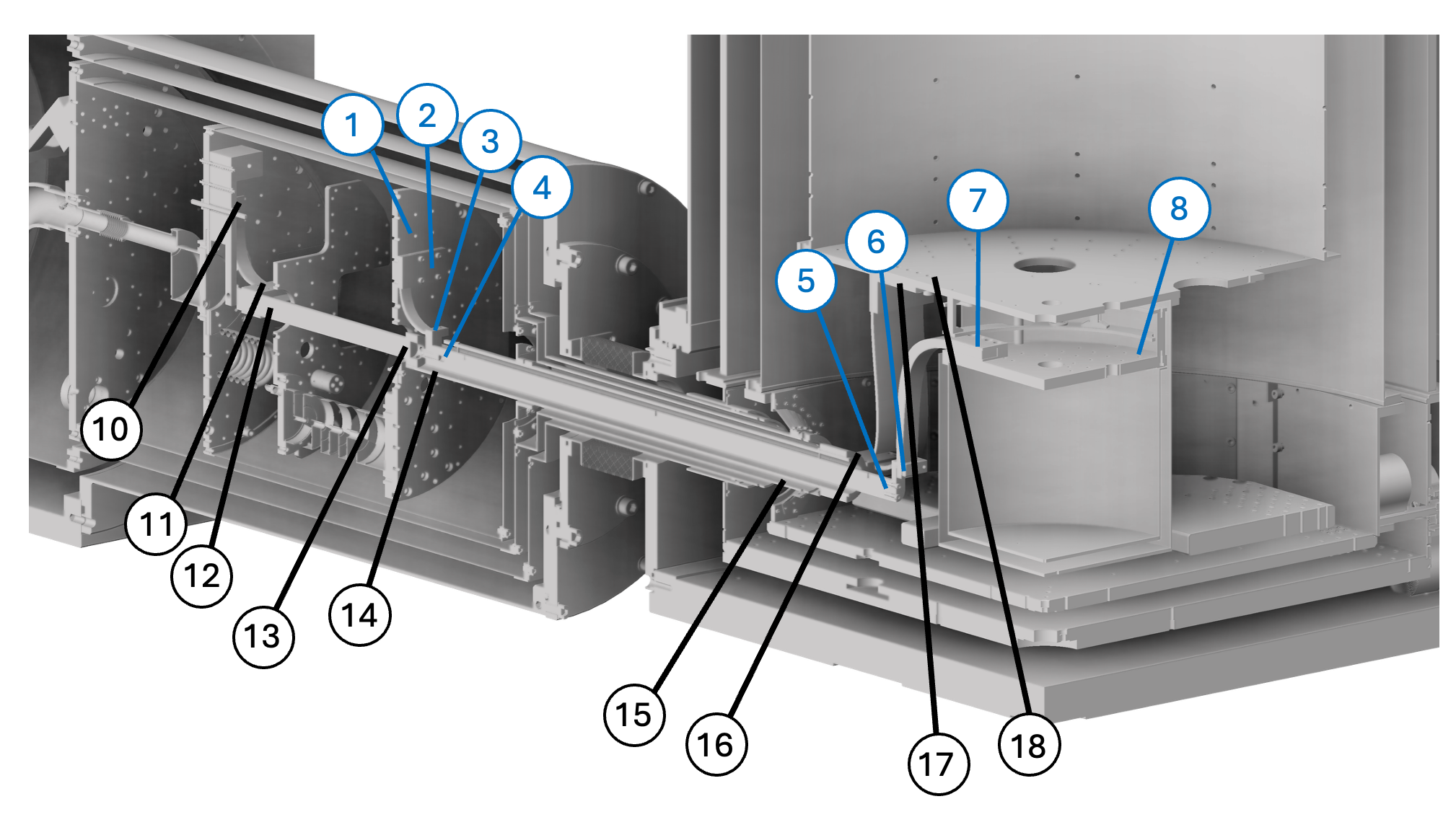} 
    \caption{CAD image with labels corresponding to the locations of the measurements shown in Figure \ref{fig:heater_measurements}.}
    \label{fig:heater_measurements_cad}
\end{figure}

\begin{figure}[h]
    \centering
    \includegraphics[width=1\textwidth]{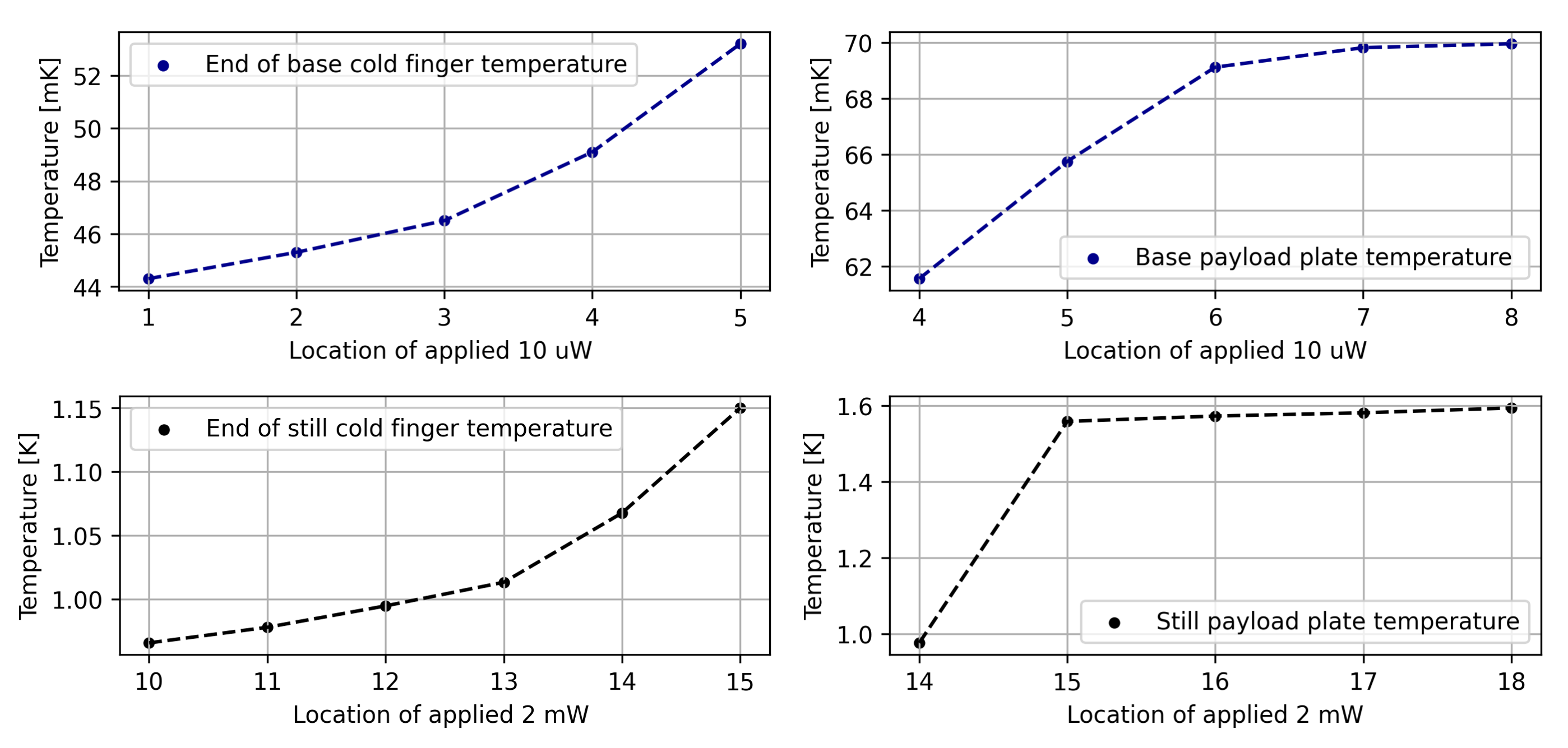} \hspace{1cm}
    \caption{Plots of thermal measurements of the base path (top) and still path (bottom). The locations represented by numbers are shown in Figure~\ref{fig:heater_measurements_cad}. Each plot corresponds to an individual configuration, so while the absolute temperatures are different, the relative differences within each plot show that the cold fingers always have the largest jump in temperature. Locations 4 and 5 are either side of the base cold finger, and locations 14 and 15 are either side of the still cold finger section going through the tight constriction between the DR and payload cryostat volumes.}
    \label{fig:heater_measurements}
\end{figure}

The thermal conductances along the still and base stage thermal paths were determined by applying calibrated heater power (\SI{2}{\milli\watt} and \SI{10}{\micro\watt}, respectively) to strategically placed heaters and measuring the resulting temperature. These measurements for both the base stage and still stage are shown in Figure~\ref{fig:heater_measurements}, and the numbers representing the locations of the applied heat are shown in Figure~\ref{fig:heater_measurements_cad}. For each segment, the slope was used to extract the effective thermal conductance. Importantly, even a qualitative interpretation of these measurements demonstrates that the dominant temperature drops occur across the long cold fingers, as expected in our design. 


With the cold fingers identified as the primary limiting factor, several avenues exist for future improvements to the thermal paths. The cold fingers could be fabricated from higher-purity Cu or annealed to increase the residual resistivity ratio. Furthermore, their geometry could be optimized to achieve a larger area-to-length ratio, enhancing thermal conductance.

Overall, the measured steady-state temperatures, thermal path conductance, and inferred heat loads are in good agreement with the thermal model and meet or exceed the requirements outlined in Section~\ref{sec:reqs}. The system provides adequate cooling power at all stages, validating the effectiveness of the flexible thermal links, radiation shielding, and low-conductivity mechanical supports. These results demonstrate that the cryogenic platform is capable of sustained mK operation suitable for low-noise resonant axion searches.

\section{Outlook and Conclusion}
\label{sec:conclusion}

We have designed, constructed, and commissioned the cryogenic system for the DMRadio-50L axion dark matter experiment. The system combines a horizontal DR with a vertically-oriented payload cryostat to simultaneously satisfy several requirements: cooling a large experimental volume across multiple temperature stages, supporting a mechanical load approaching \SI{200}{\kilo\gram}, and maintaining mK temperatures with sufficient cooling power for the payload. The system was also designed to operate within a  laboratory environment with conventional ceiling height constraints. Future work on the DMRadio-50L experiment will focus on full detector integration and extended physics data-taking.

The hybrid architecture successfully decouples mechanical support and thermal transport functions. Flexible connections such as Cu straps, compliant Al foils, and concentric cylinders with fingerstrips enable high conductance while accommodating differential thermal contraction. Low-conductance mechanical connections such as the Vespel screw suspensions, Vespel compression support, Kevlar puck, and G10 tubes provide mechanical support throughout the thermal paths without introducing significant heat loads. 

Commissioning measurements demonstrate that all temperature stages meet or exceed their design specifications. The pulse tubes provide ample cooling power at \SI{40}{\kelvin} and \SI{4}{\kelvin} for radiation shielding and magnet operation, the still stage provides an effective thermal intercept for superconducting shielding between 500 - \SI{700}{\milli\kelvin}, and the MXC achieves base temperatures near \SI{20}{\milli\kelvin}. The base payload plate achieves temperatures below \SI{50}{\milli\kelvin}, allowing sensitive detector operation. Measured heat loads and conductances are consistent with the thermal model. 

Potential improvements focus primarily on further reducing parasitic heat loads and lowering the effective thermal impedance between the DR and the detector payload. Although the present system meets all operational requirements, measurements indicate that residual radiative loading and finite conductance of several long thermal links remain the dominant contributors to the temperature margin at the still and base stages. Additional radiation shielding could therefore provide immediate benefits. In particular, increasing the coverage of surfaces painted black, extending light-tight baffling around shield interfaces, and ensuring adequate shielding and routing of hotter wiring would further suppress direct line-of-sight radiation and reduce the photon-mediated heat load from warmer surfaces. 

Further improvements can be realized by upgrading the cold fingers themselves. Replacing existing C101 Cu components with higher-purity or annealed Cu with larger residual resistivity ratio would increase low-temperature thermal conductivity and thereby reduce longitudinal temperature gradients. Because the cold fingers dominate the total thermal resistance of both the still and base cooling paths, even modest improvements in conductivity translate directly into measurable reductions in payload temperature and increased available cooling power. Similar improvements could be obtained by optimizing cross-sectional geometry or adding parallel conduction paths where mechanical constraints permit.

Other elements of the thermal chain may also be improved. The flexible Cu straps and mechanical interfaces represent additional series resistances that can be reduced through increased foil count, higher clamping forces, or improved Au-plated contact surfaces to minimize boundary resistance. Redesigning certain joints to increase effective contact area or to incorporate welded connections would further enhance conductance while preserving the mechanical compliance required to accommodate differential contraction. 

Finally, continued refinement of vibration isolation will likely become more important as sensitive detector components are integrated. Reducing mechanical coupling to the payload will improve detector noise performance. The payload cryostat is designed with four flanges in the S2 payload plate, S1 payload plate, and octagonal base chamber that would allow the installation of alternative supports for the S2 payload ring. Since the payload is fully supported by the S2 payload ring and is mechanically isolated from the DR and the pulse-tube cryocooler, the alternative supports would isolate the payload from several possible sources of vibrations. 

Together, these incremental improvements provide a clear path toward lower base temperatures, increased cooling margin, and expanded capacity for future detector components without requiring substantial changes to the overall cryostat architecture.

Beyond enabling the DMRadio-50L science program, this system provides a practical implementation of DR-based cooling to large detector masses and volumes. The modular interface between the two cryostats allows independent testing and maintenance, while the compliant thermal connections and staged radiation shielding provide a generalizable solution for large experiments requiring mK cooling. The same features that make the platform suitable for the axion search, a stable, low bath temperature, ample volume for test circuitry, strong magnetic field screening, and access to a high-$Q$ superconducting resonator in the \si{\mega\hertz} range, also make it well suited as a testbed for quantum sensors such as the RQU~\cite{Kue2024}. By supporting the development of readout below the standard quantum limit, the platform will help enable faster axion searches in addition to advancing quantum sensing more broadly.

\acknowledgments
DMRadio-50L is supported by the Gordon and Betty Moore Foundation Grant No. 7941 and the Heising-Simons Foundation Grant No. 128583. Further support is provided by DOE and NSF grants to individual institutions. MIT's contribution is supported by the National Science Foundation under Grant No. 2411650. Chelsea Bartram and Andrew Yi are supported by the Department of Energy, Laboratory Directed Research and Development program at SLAC National Accelerator Laboratory, under contract DE-AC02-76SF00515 and as part of the Panofsky Fellowship awarded to Chelsea Bartram. Pamela Stark is supported by the National Science Foundation Graduate Research Fellowship under Grant No. DGE-2146755. J. T. Fry is supported by the National Science Foundation Graduate Research Fellowship under Grant No. 2141064. The authors thank Maria Salatino for valuable early discussions and technical insight. The authors also thank the MIT T.J. Rodgers RLE Laboratory for their helpful equipment and advice.

\newpage
\appendix
\section{Dimensions and Interfaces}
\label{appendix:dimensions}
\begin{table}[h]
\centering
\caption{Dimensions and materials of cryogenic structural and thermal components.}
\label{table:dimensions_all}
\scriptsize
\begin{tabular}{>{\centering\arraybackslash}p{3.2cm}
                >{\centering\arraybackslash}p{1.1cm}
                >{\centering\arraybackslash}p{1.1cm}
                >{\centering\arraybackslash}p{1.1cm}
                >{\centering\arraybackslash}p{1.1cm}
                >{\centering\arraybackslash}p{1.1cm}
                >{\centering\arraybackslash}p{1.4cm}
                >{\centering\arraybackslash}p{1.9cm}}
\hline
\textbf{Component} & \textbf{Stage} & \textbf{Length / Height (cm)} & \textbf{Width (cm)} & \textbf{Inner Diameter (cm)} & \textbf{Outer Diameter (cm)} & \textbf{Thickness (mm)} & \textbf{Material} \\
\hline
\multicolumn{8}{l}{\textit{Vacuum chamber --- DR}} \\
Cubic chamber& Vacuum & 54.7 / 54.7  & 48.3 & — & — & 25 & Al \\
Vacuum shield & Vacuum & 64.1 & — & 41.53 & 42.17 & 3.2 & Al \\
Vacuum shield bottom & Vacuum & — & — & — & 42.17 & 20 & Al \\

\hline
\multicolumn{8}{l}{\textit{Vacuum chamber --- payload cryostat}} \\
Octagonal chamber & Vacuum & 21.9 & — & — & 10.41 & 38 & Al6061 \\
Vacuum chamber lid & Vacuum & 90 & — & 100 & 101 & 5 & Al6061 \\
\hline
\multicolumn{8}{l}{\textit{First stage --- DR}} \\
1st stage DR plate & PT 1st & — & — & — & 44.7 & 12 & Cu (+ Au) \\
Radiation shield 1 (wide) & PT 1st & 28.5 & — & 40.8 & 41 & 1 & Al \\
Radiation shield 1 (narrow) & PT 1st & 64.5 & — & 36.4 & 36.6 & 1 & Al \\
Shield 1 bottom & PT 1st & — & — & — & 36.5 & 2 & Al6063 \\
Radiation shield 2 (wide) & PT 1st & 2.4 & — & 13.3 & 13.6 & 1.5 & Al6063 \\
Radiation shield 2 (narrow) & PT 1st & 30.8 & — & 6 & 6.4 & 2 & Al6063 \\

\hline
\multicolumn{8}{l}{\textit{First stage --- payload cryostat}} \\
1st stage payload plate & PT 1st & — & 90 & — & — & 12 & C101 Cu \\
Radiation shield 3 & PT 1st & 85 & — & 90 & 90.6 & 3 & Al6061 \\
Concentric cylinder & PT 1st & 8.7 & — & 6.82 & 7.22 & 2 & Al7075 \\

\hline
\multicolumn{8}{l}{\textit{Second stage --- DR}} \\
2nd stage DR plate & PT 2nd & — & — & — & 35.2 & 10 & Cu (+ Au) \\
Radiation shield 1 & PT 2nd & 62 & — & 32 & 32.2 & 1 & Al \\
Shield 1 bottom & PT 2nd & — & — & — & 32.5 & 2 & Al6063 \\
Radiation shield 2 (wide) & PT 2nd & 3.1 & — & 11.6 & 12 & 2 & Al6063 \\
Radiation shield 2 (narrow) & PT 2nd & 38.25 & — & 4.6 & 5 & 2 & Al6063 \\

\hline
\multicolumn{8}{l}{\textit{Second stage --- payload cryostat}} \\
2nd stage payload plate & PT 2nd & — & 78.5 & — & — & 12 & C101 Cu \\
Radiation shield 3 & PT 2nd & 83.8 & — & 80 & 80.3 & 1.5 & C110 Cu \\
Al foil strips & PT 2nd & 4.4 & 1.5 & — & — & 0.05 & 99.99$\%$ Al \\

\hline
\multicolumn{8}{l}{\textit{Still stage}} \\
Still plate & Still & — & — & — & 30.5 & 9 & Cu (+ Au) \\
Thermal strap 1 foils & Still & 5 & 2 & — & — & 28 $\times$ 0.127 & C110 Cu \\
Cold finger section 1 (rod) & Still & 26 & — & — & 2 & — & C101 Cu \\
Cold finger section 2 (tube) & Still & 63 & — & 3 & 3.6 & 3 & C101 Cu \\
Thermal strap 2 foils & Still & 20 & 12 & — & — & 29 $\times$ 0.127 & C110 Cu\\
Radiation shield 1 & Still & 41.2 & — & 30.3 & 30.4 & 0.5 & Cu \\
Shield 1 bottom & Still & —  & — & — & 30 & 2 & C101 Cu \\
Still payload plate (top) & Still & — & — & — & 70 & 5.85 & Al6061 \\
Still payload plate (bottom) & Still & — & — & — & 70 & 1 & C101 Cu \\
Radiation shield 2 & Still & 23.5 & — & 29.8 & 30 & 1 & Al6061 \\
Radiation shield 3 & Still & 66 & — & 66 & 66.6 & 3 & Al1100 \\
G10 support tubes & Still & 22.3 & — & 3.02 & 3.32 & .8 & G10 \\

\hline
\multicolumn{8}{l}{\textit{Base stage}} \\
MXC plate & Base & — & — & — & 29 & 8 & Cu (+ Au) \\
Base thermal strap 1 foils & Base & 5 & 2 & — & — & 20 $\times$ 0.127 & C110 Cu \\
Base cold finger & Base & 65 & — & — & 2 & — & C101 Cu (+ Au) \\
Base thermal strap 2 foils & Base & 20 & 3.5 & — & — & 62 $\times$ 0.127 & C110 Cu \\
Base payload plate & Base & — & — & — & 28.9 & 10 & C101 Cu (+ Au) \\
Base stage radiation shield & Base & 16 & — & 27.1 & 27.3 & 1 & Al6061 \\
Vespel support screws & Base & 0.8 & — & — & M2 & — & Vespel SP-1 \\
Vespel dumbbells & Base & 2.28 & — & 0.1875 & 0.4 & — & Vespel SP-1 \\
\hline
\end{tabular}
\end{table}
\begin{table}
\centering
\caption{Thermal interface parameters along primary conductive pathway for still and base stages. All screws listed here are Al7075, and torque values are provided where it is practical to use a torque wrench.}
\label{table:interfaces}
\scriptsize
\begin{tabular}{>{\centering\arraybackslash}p{3.8cm}
                >{\centering\arraybackslash}p{1.3cm}
                >{\centering\arraybackslash}p{1.5cm}
                >{\centering\arraybackslash}p{1.3cm}
                >{\centering\arraybackslash}p{1.3cm}
                >{\centering\arraybackslash}p{1.3cm}
                >{\centering\arraybackslash}p{2cm}}
\hline
\textbf{Interface} & \textbf{Stage} & \textbf{Interface Material} & \textbf{No. of Screws} & \textbf{Screw Size} & \textbf{Torque (in·lbs)} & \textbf{Interface Treatment} \\
\hline
Still plate → still thermal strap 1 & Still & Au–Cu & 8 & M3 & 5 & — \\
Still thermal strap 1 → still cold finger rod (round clamp) & Still & Cu–Cu & 8 & M3 & — & Scotch-Brite + Apiezon N grease \\
Still cold finger rod → still cold finger tube & Still & Cu–Cu & 4 & M4 & 16 & Scotch-Brite + Apiezon N grease \\
Still cold finger tube → still thermal strap 2 (round clamp) & Still & Cu–Cu & 8 & M4 & 8 & Scotch-Brite + Apiezon N grease \\
Still thermal strap 2 → still payload plate & Still & Au–Au & 6 & M6 & 28 & — \\

MXC plate → base thermal strap 1 & Base & Au–Au & 8 & M3 & 5 & — \\
Base thermal strap 1 → base cold finger & Base & Au–Au & 1 & M5 & 25 & — \\
Base cold finger → base thermal strap 2 & Base & Au–Au & 2 & M3 & — & — \\
Base thermal strap 2 → base payload plate & Base & Au–Au & 10 & M4 & 12 & — \\

\hline
\end{tabular}
\end{table}


\newpage
\bibliographystyle{JHEP}
\bibliography{biblio.bib}



%


\end{document}